\documentclass[twocolumn]{aastex63}
\hypersetup{linkcolor=red, citecolor=blue}
\usepackage{comment}
\usepackage{amssymb}
\usepackage{amsfonts}
\usepackage{amsmath}
\usepackage[varg]{txfonts}
\usepackage{natbib}
\usepackage{color}
\usepackage{graphicx}
\usepackage{hyperref}

\newcommand{\RNum}[1]{\uppercase\expandafter{\romannumeral #1\relax}}
\newcommand{\felix}[1]{\textsc{F\small{ELIX}}}

\newcommand\aastex{AAS\TeX}

\received{---}
\revised{---}
\accepted{---}
\submitjournal{ApJ}

\shorttitle{\aastex\ Mode behaviours in the sdB and white dwarf stars}
\shortauthors{Zong et al. 2020}
\begin{document}

\title{Oscillation mode variability in evolved compact pulsators from {\sl Kepler} photometry. \RNum{2}. Comparison of modulation patterns between raw and corrected flux}

\email{weikai.zong@bnu.edu.cn}

\author{Weikai Zong}
\affil{Department of Astronomy, Beijing Normal University, Beijing~100875, P.~R.~China}

\author{St\'ephane Charpinet}
\affil{IRAP, Universit\'e de Toulouse, CNRS, UPS, CNES, 14 avenue Edouard Belin, F-31400, Toulouse, France}

% \author{Jian-Ning Fu}
% \affil{Department of Astronomy, Beijing Normal University, Beijing~100875, P.~R.~China}

\author{G\'erard Vauclair}
\affil{IRAP, Universit\'e de Toulouse, CNRS, UPS, CNES, 14 avenue Edouard Belin, F-31400, Toulouse, France}


%%\Large
\begin{abstract}

We present the second results of an ensemble and systematic survey of
oscillation mode variability in compact pulsators observed with the 
original {\sl Kepler} mission. Two types of flux calibrations, raw and corrected, collected on two 
hot B subdwarf stars, KIC\,2438324 and KIC\,11179657, are thoroughly examined
with the goal to evaluate the difference of patterns when oscillation modes modulate in 
amplitude (AM) and frequency (FM). We concentrate on AMs and FMs occurring in seven 
multiplet components in each star as representative frequencies. The analysis shows 
that FM measurements are independent of the flux calibration we choose. 
However, if flux 
contamination by nearby stars is large, AMs may be significantly different between raw and corrected flux. 
In addition, AMs suffer, to some extent, from systematic modulation pattern {which is} 
most likely induced by instrumental effects {and} differs from one star to another. 
%Fortunately, these can be corrected {through careful inspection} with the help of reference signals. 
Our results indicate that stars with no contamination are better candidates to quantitatively compare modulation patterns
with theory and should be given a higher priority for such studies, since {light contamination} will destroy real {amplitude modulation patterns}. 

\end{abstract}

\keywords{technique: photometric --- stars: evolved variables --- stars: oscillations}



\section{Introduction}
This series of papers is devoted to an ensemble and systematic survey of oscillation 
mode variability in pulsating hot B subdwarf (sdB) and white dwarf stars using {\sl Kepler}'s photometry, 
particularly for stars with over two years of monitoring. We recall, in the first 
paper of this series \citep[][hereafter Paper\,\RNum{1}]{zong2018}, that a particular sdB star, 
KIC\,3527751, was thoroughly analyzed with a total of 204 frequencies resolved from a nearly 
contiguous 38-month long light curve. We investigated mode variability and stability for 143 frequencies 
with relatively large amplitudes (i.e., {with acceptable significance level in just a portion} of 
the entire light curve) and found that all of those frequencies show evident variations, with regular 
or irregular modulation patterns. We showed that these variations are likely reminiscent of nonlinear weak mode interactions predicted by the resonant mode coupling formalism \citep[e.g.,][]{buchler1997}, where 
the oscillation mode under certain resonance conditions may exhibit temporal amplitude modulations 
(AM) and frequency modulations (FM) of various patterns or stay stable over time. Such observed 
modulation patterns will eventually be compared to theoretical ones, once calculations of nonlinear coupling coefficients involving many modes \citep{buchler1995} will become available. The direct comparison of AMs and FMs has only been achieved for a few pulsating stars \citep[see, e.g., ][]{kovacs1989}, thus far. Therefore, the current and most urgent step is to provide intrinsic AM and FM measurements as this opportunity is now offered by {\sl Kepler}. Recent results show 
that pulsating sdB and white dwarf stars may be among the best candidates on this front 
\citep[see, e.g.,][]{zong2016a}.

In the orginal {\sl Kepler} field, the satellite collected exquisite high-quality photometry 
for eighteen pulsating sdB stars and six white dwarf stars (see Paper\,\RNum{1}, and references therein). 
Most of these objects have been intensively observed in the competitive short-cadence (58.85\,s) mode over 
a duration of two years due to their rapid oscillation timescales (from a few minutes to a few hours) 
and scientific significance. {\sl Kepler} pipelines provide available light curves in two forms: 
the ``raw'' flux, also referred to as the Simple Aperture Photometry (SAP), and the ``corrected'' flux, also called Pre-search Data Conditioning SAP \citep[PDCSAP; see, e.g.,][]{2010ApJ...713L..87J}. The latter data impose cotrending basis vectors to correct the discontinuties and contamination of raw flux over 
different quarters. Although the corrected data contain cleaner light curves, this process may bring extra 
or modify intrinsic astrophysical signatures of some particular targets 
\citep[see][]{2012MNRAS.422..665M}. In the context of precisely characterizing modulation 
patterns in oscillation modes, the impact of using two different types of flux has never been examined. Concretely, evaluating these differences for some particular stars might be necessary before attempting quantitative comparisons with theoretical calculations.

With this purpose in mind, we chose two long-period {\sl g}-mode pulsating sdB stars, KIC\,2438324 and KIC\,11179657, as representative objects to check if using two different kinds of fluxes leads to
significant differences in the modulation patterns. 
KIC\,2438324, or B4 in NGC\,6791, has a mean brightness {\sl Kp} = 18.267, effective temperature 
$T_\mathrm{eff} = 24\,786 \pm 655$\,K and surface gravity $\log g = 5.30 \pm 0.09$\,dex 
\citep{2012MNRAS.427.1245R}. The detection of pulsations in this very faint star required the acquisition of over six months of {\sl Kepler} photometry \citep{2011ApJ...740L..47P}, which established its variability.   
KIC\,11179657, or USNO-A2.0 1350-10140904, has a mean brightness {\sl Kp} = 17.065, 
$T_\mathrm{eff} = 26\,000 \pm 800$\,K and $\log g = 5.14 \pm 0.13$\,dex \citep{2010MNRAS.409.1470O}.
For that star, the presence of pulsations was revealed during the first year survey phase, on the basis of a $\sim30$\,days light curve \citep{2010MNRAS.409.1470O}. 
Both stars are found to be members of a binary system with low-mass main-sequence companions, and both are not synchronized, as suggested by the seismic rotation periods which are longer than the orbital periods \citep{2011ApJ...740L..47P,2012MNRAS.422.1343P}. 

In this paper, we first assess the robustness of our error estimates for the amplitude, frequency and phase from quantitative tests in Section~2. Section~3 is dedicated to the thorough analysis of the {\sl Kepler} photometry collected for these two sdB stars. The extracted modulation patterns for representative mode frequencies and comparisons with the orbital signals are presented in Section~4 and 5, followed by a discussion in Section~6 and a conclusion.

\section{Testing the robustness of error estimates} \label{s:r}
Considering the importance of error estimates for our scientific goals, we made a series of simulations to assess their robustness for frequencies extracted from the {\sl Kepler} photometry.
These tests constitute an extension of those described in \citep{zong2016b}. A variant of such simulations was also carried out recently by \citet{2018A&A...611A..85S}, in order to check the reliability of errors for measured frequencies that suggest the presence of a giant planet around V391\,Peg. 

\begin{figure*}
\centering
 %\epsscale{.80}
\includegraphics[width=5.6cm]{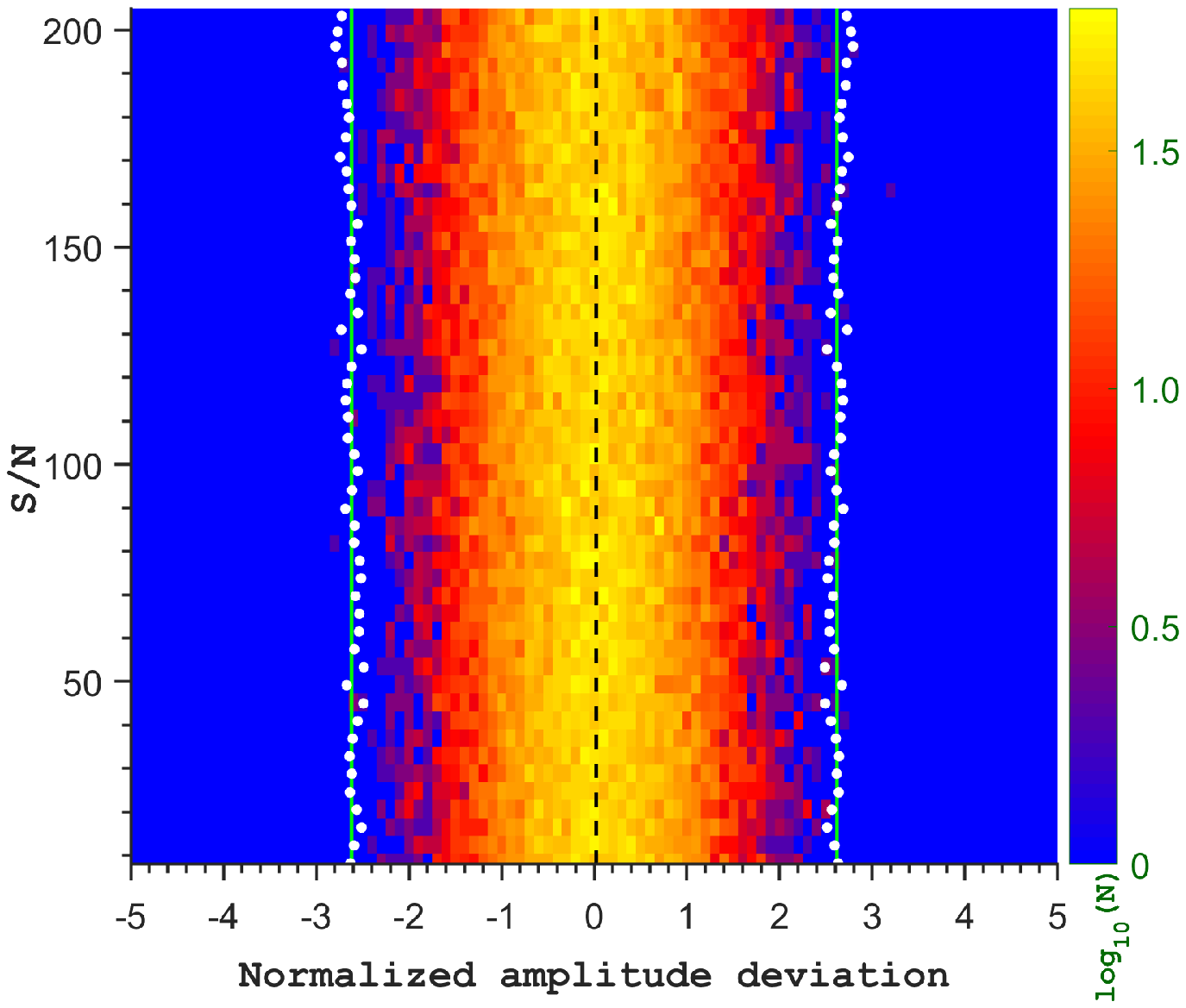}
\includegraphics[width=5.6cm]{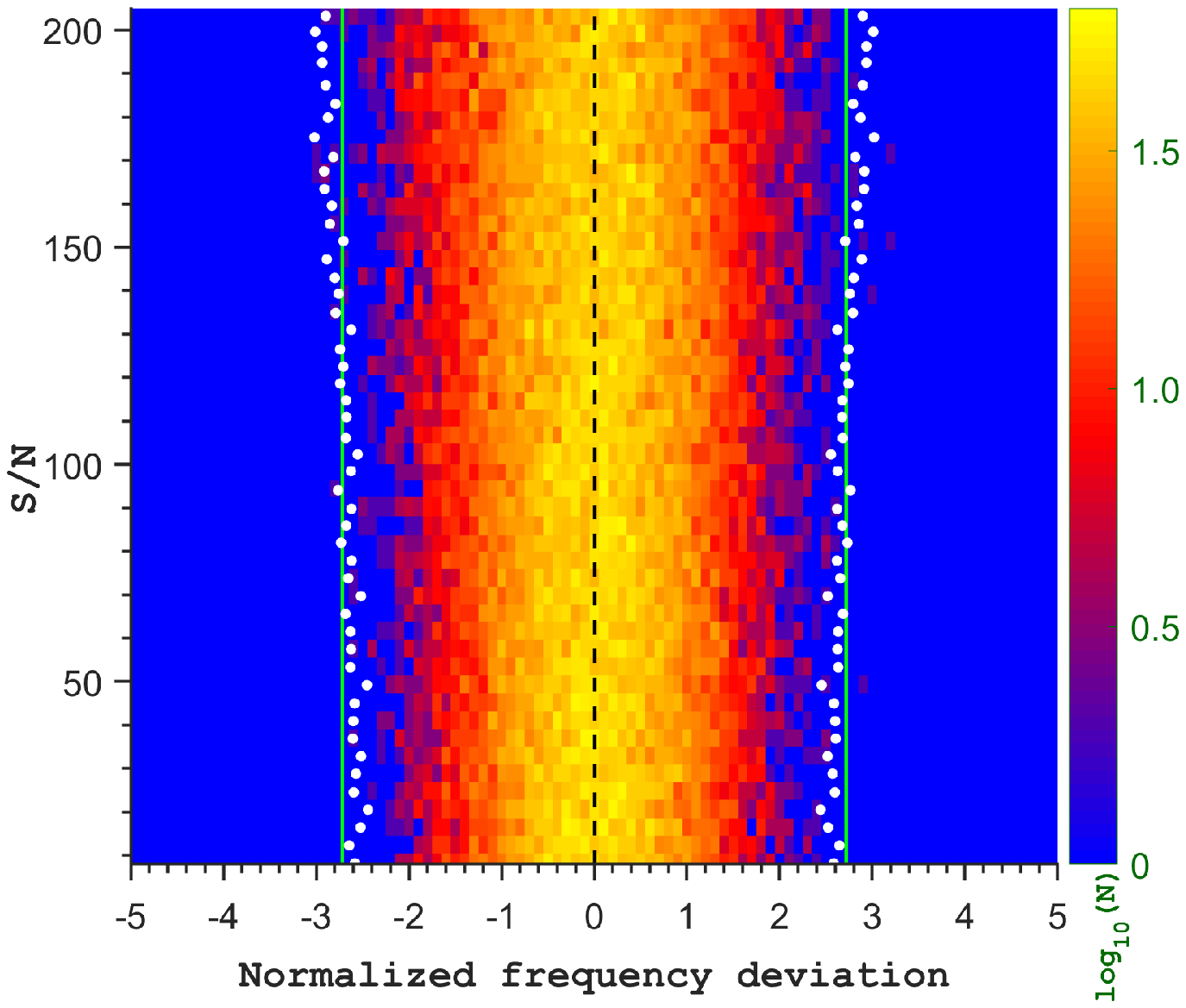}
\includegraphics[width=5.6cm]{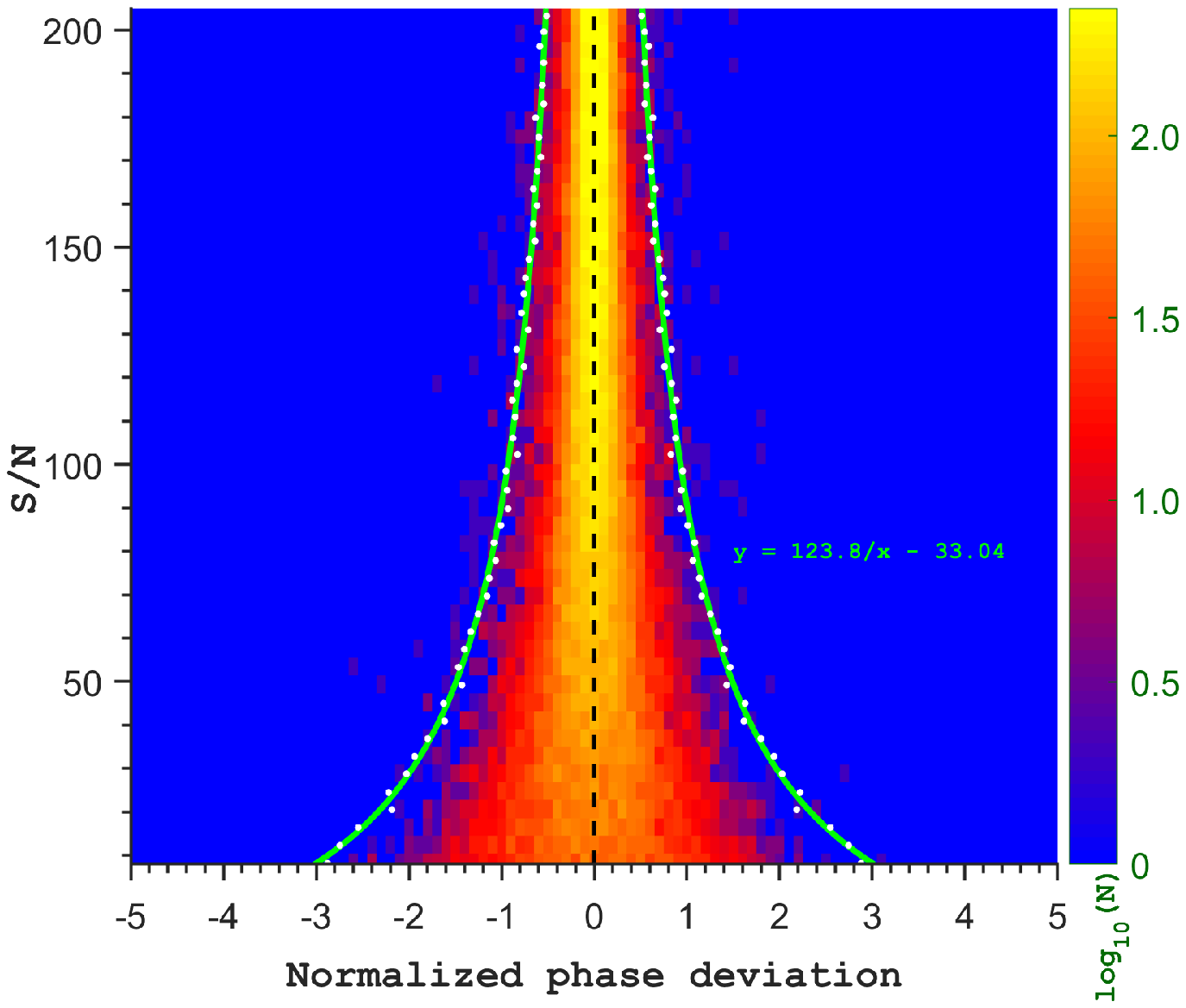}\\
\includegraphics[width=5.6cm]{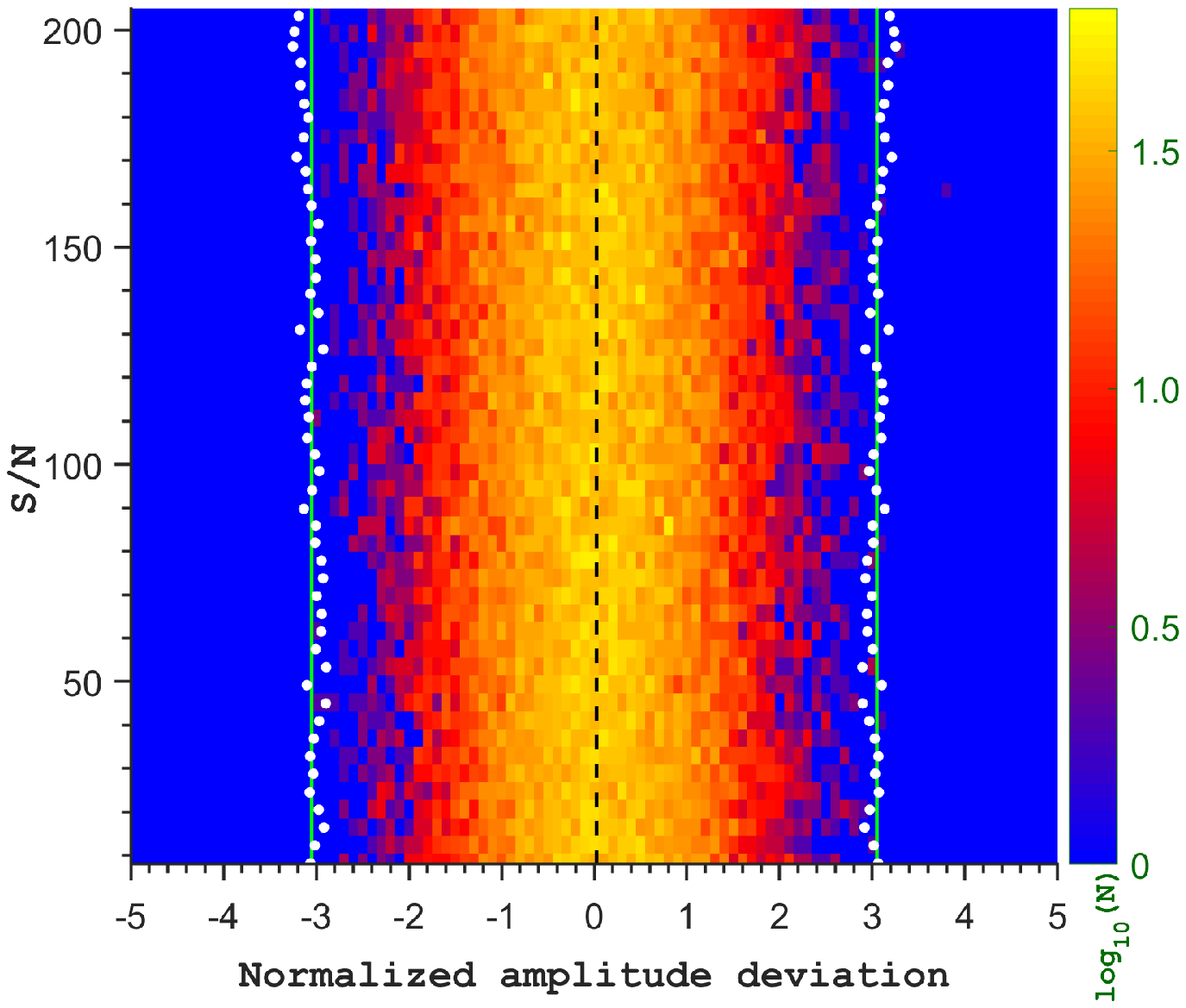}
\includegraphics[width=5.6cm]{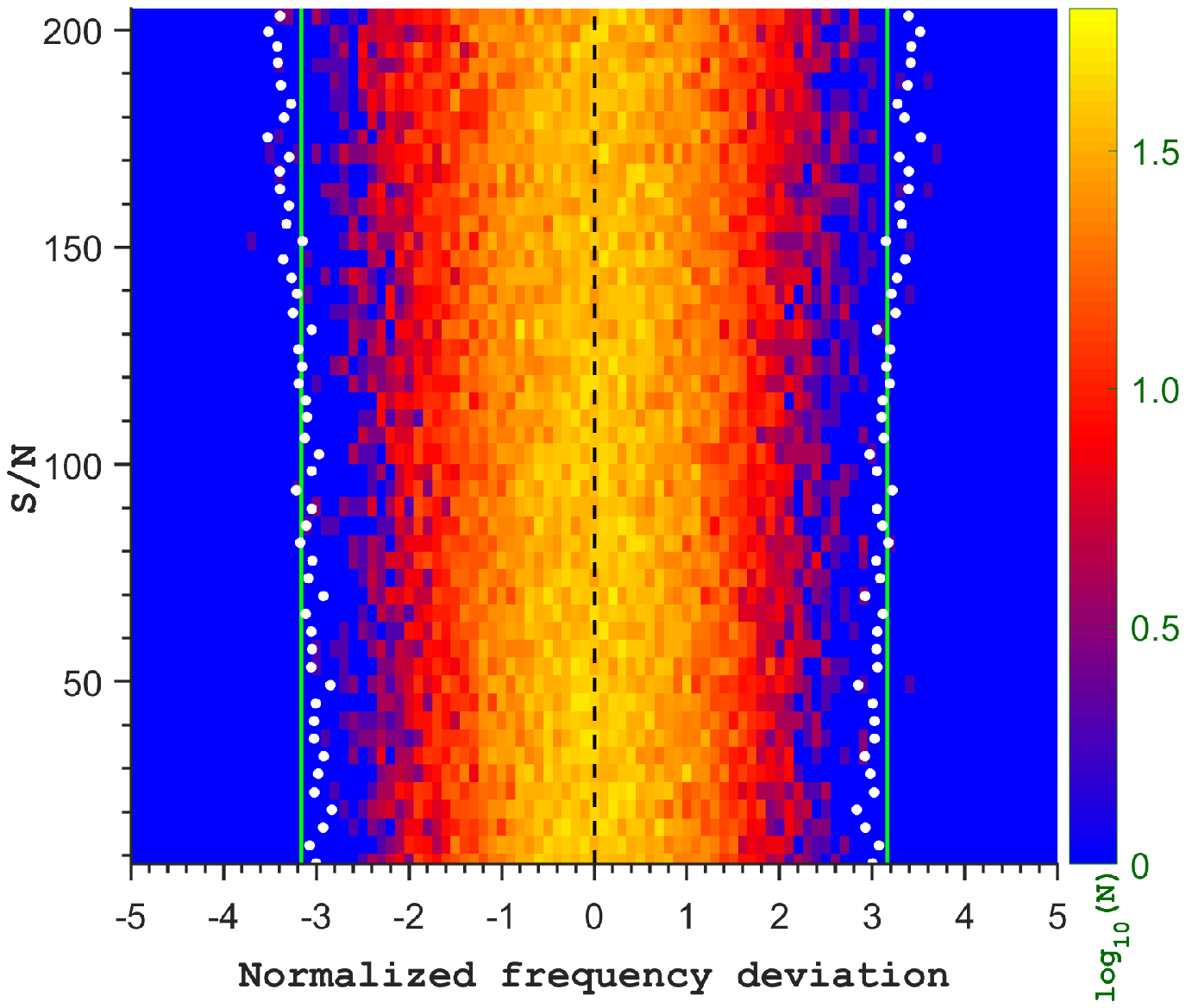}
\includegraphics[width=5.6cm]{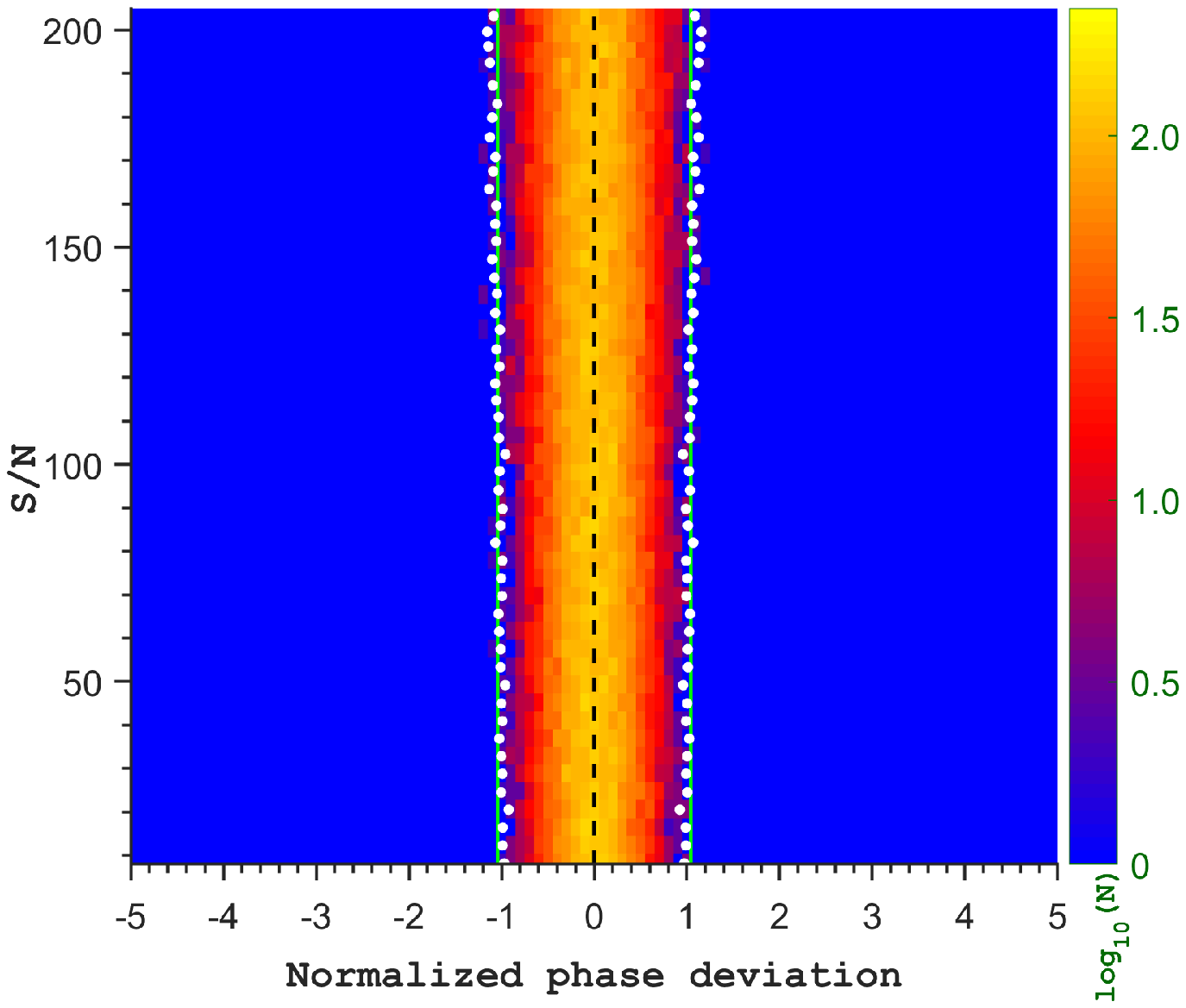}
\caption{Distribution of normalized amplitude, frequency and phase deviations between the extracted and injected signals (from left to right panels). Top panels correspond to the default method to compute errors used with the code \felix{} (see text) and bottom panels correspond to errors determined analytically following \citet{1999DSSN...13...28M}. The (central) dashed lines indicates the mean deviations calculated from the distribution, which are all very close to zero. The dotted curves indicate $\pm3\times$ the standard deviations as a function of S/N.  In all panels except the top-right one, S/N-averaged values for $3\sigma$ are represented by vertical solid lines. For the top-right panel, the solid curve is fitted using a reciprocal model. Color coding refers to the occurrence (using a logarithmic scale) of measurements at specifics deviation and S/N values.
\label{simulation}}
\end{figure*}

We briefly describe how this series of simulations was done: 1) 50 artificial light curves containing Gaussian white noise were generated using a time sampling of 58.85\,s and a duration of 200 days without interruption; 2) in each light curve, 1\,000 frequencies of constant amplitude \citep[details are provided in][]{zong2016b} were injected, with amplitude values varying from light curve to light curve; 3) The code \felix{}\footnote{Frequency Extraction for LIghtcurve eXploitation \citep[see details in][]{charpinet2010,charpinet2011,zong2016b}} was used to detect and extract automatically the injected frequencies in each simulated light curve; 4) two estimates for the normalized errors were calculated, using different methods. 
One method was to derive analytically the errors following \citet[][hereafter MO99]{1999DSSN...13...28M}, with the relations for amplitude, frequency and phase uncertainties : 
\begin{subequations} \label{am111}
\begin{alignat}{3}
  \sigma_\mathrm{A} & = \sqrt{2/N}\,\sigma_\mathrm{m},  \label{seq61} \\              
  \sigma_\mathrm{f} & = \frac{\sqrt{3}}{\pi\,T} \frac{\sigma_\mathrm{A}}{A},  \label{seq62} \\     
  \sigma_\mathrm{\phi} & = \sigma_\mathrm{A} / A,  \label{seq63}   
\end{alignat}
\end{subequations}
where $N$ and $T$ are the number of data points and the total duration of the time-series, respectively. $A$ and $\sigma_\mathrm{m}$ are the measurable amplitude of a frequency peak and the root-mean-square deviation of the magnitude in the light curve, respectively. The other method which was used thus far in the code \felix{} directly measures $\sigma_\mathrm{A}$ as the median value of the noise around a detected peak in the Fourier transform of the light curve. $\sigma_\mathrm{f}$ is then calculated with equation\,\ref{seq62}) and $\sigma_\mathrm{\phi}$ comes from the covariance matrix of the nonlinear least-square fit of the light curve. The normalized errors for the amplitude, frequency and phase are defined as  
\begin{subequations} \label{am111}
\begin{alignat}{3}
  \Delta_\mathrm{A} & = (A_\mathrm{pre} - A_\mathrm{inj}) /  \sigma_\mathrm{A}, \label{seq71} \\              
  \Delta_\mathrm{f} & = (f_\mathrm{pre} - f_\mathrm{inj}) /  \sigma_\mathrm{f}, \label{seq72} \\    
  \Delta_\mathrm{\phi} & = (\phi_\mathrm{pre} - \phi_\mathrm{inj}) /  \sigma_\mathrm{\phi}. \label{seq73}  
\end{alignat}
\end{subequations}
respectively. Subscripts ``pre'' and ``inj'' refer to prewhitened (measured) and injected values.

Figure\,\ref{simulation} shows the results of our simulations from a total of 50\,000 injected sinusoidal waves. By using normalized errors, one expects to obtain normal distributions of zero mean and standard deviation one, {$N\sim(0,1)$}, if the values of $\sigma_\mathrm{A}$, $\sigma_\mathrm{f}$, and $\sigma_\mathrm{\phi}$ are correctly estimated. {A narrower distribution than $N\sim(0,1)$ gives an overestimation of error.}
We clearly see that all prewhitened values are very similar to the injected ones, as revealed by the vertical dashed lines near the center. This indicates that there is no bias in the frequency extraction method. Amplitude and frequency uncertainties evaluated from MO99 equations and from our implemented method are both found to be in agreement with the normal distribution in general. However, ours slightly overestimates $\sigma_\mathrm{A}$ and $\sigma_\mathrm{f}$ by about 10$\%$ and 5$\%$, while MO99 slightly underestimate these errors by about 2$\%$ and 5$\%$, respectively. 
We note that there is possibly a slight dependence to S/N of the normalized frequency deviations (i.e., the dashed lines at the low and high S/N ends stand at two sides of the solid vertical lines). 
The phase deviations, however, show two different behaviors. The one derived from the covariance matrix of the nonlinear fit presents a clear dependence on S/N as it become gradually overestimated with increasing signal-to-noise ratio (or amplitude). For its part, MO99 gives a constant overestimate by a factor of about three for all S/N ratios. We note that Equation\,\ref{seq63} is obtained by averaging the observed time. A change in the zero-point of time will introduce a constant factor to that equation that could correct for the observed difference. While investigating further the reasons behind these differences could be of interest, we stick to our main goal here, which is to calibrate our method to estimate errors accurately. Based on the tests presented above, we thus validated our method to compute uncertainties for the amplitude and frequency, and we implemented a normalized method (MO99 formula divided by a constant value 3) to estimate phase uncertainties.

\section{{\sl Kepler} photometry}
%--------------------------------------------------------------------------------
\begin{deluxetable}{cccrr}
\centering
\tablecaption{Contamination values per quarter for KIC\,2438324 and KIC\,11179657 during the main campaign. \label{t1}}
\tablehead{
\colhead{Quarter} & \colhead{KIC\,2438324}& \colhead{KIC\,11179657}}
\startdata
5    &  -       &  0.0001     \\
6    & 0.046    &    0        \\
7    & 0.414    &    0        \\
8    & 0.007    &    -        \\
9    & 0.031    &  0.0002     \\
10   & 0.047    &    0        \\
11   & 0.416    &    0        \\
12   & 0.007    &    -        \\
13   & 0.032    &  0.0001     \\
14   & 0.046    &    0        \\
15   & 0.418    &    0        \\
16   & 0.007    &    -        \\
17   & 0.032    &    0        \\
\enddata
\end{deluxetable}
%==============================================================================
In this work, all data were obtained from the {\sl Kepler} space telescope during its initial survey phase. We only analyzed short-cadence (58.85\,s) light curves distributed through the MAST\footnote{The Mikulski Archive for Space Telescopes:\\ http://archive.stsci.edu/kepler/} website. This exposure allows us to detect rapid oscillations in compact pulsating stars \citep[see, e.g.,][]{charpinet2011}. KIC\,2438324 was continuously observed from Q6.1 to Q17.2, for a total duration of 35 months. KIC\,11179657 was observed in several segments, Q2.3, Q5.1-Q7.3, Q9.1-Q11.3, Q13.1-Q15.3 and Q17.1-Q17.2, because this star {fell} on the nonfunctional CCD Module 3 every four quarters, due to
the onboard photometer being rolled by 90 degrees every $\sim 93$ days. This rotation of the instrument may induce fluctuations of the contamination factor of some targets, as images of these stars relocate onto different CCD modules.
The initial one-month run Q2.3 was not further considered, due to its long disconnection with the main part of the {\sl Kepler} observations. 
Table\,\ref{t1} lists the contamination values reported at different quarters for KIC\,2438324 and KIC\,11179657. KIC\,2438324 lightcurve was contaminated by nearby stars with a factor varying 
from 0.007 to 0.418. In contrast, KIC\,11179657 lightcurve has almost never suffered pollution from the light of nearby stars, with a maximum factor of 0.0002.

\begin{figure*}
\centering
\includegraphics[width=8.5cm]{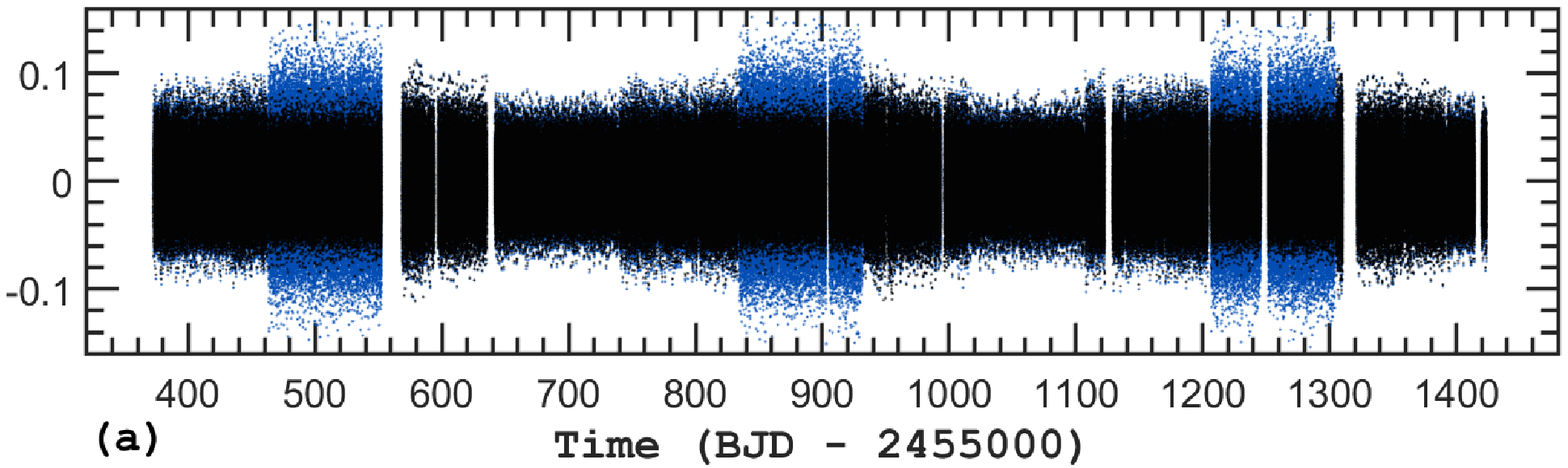}
\includegraphics[width=8.5cm]{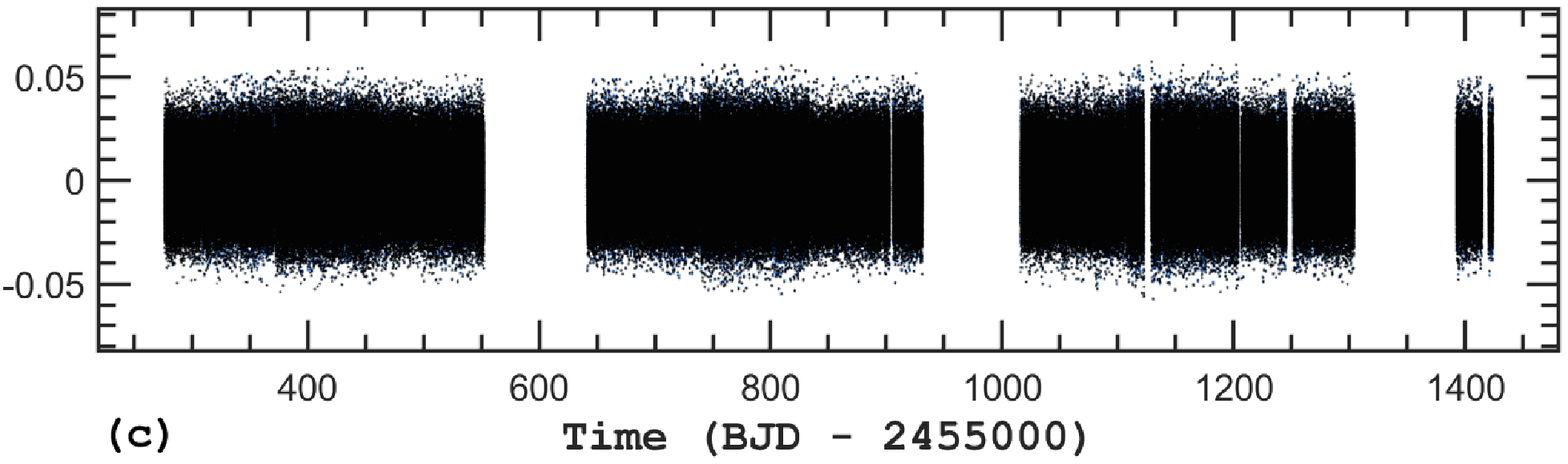}\\
\includegraphics[width=8.5cm]{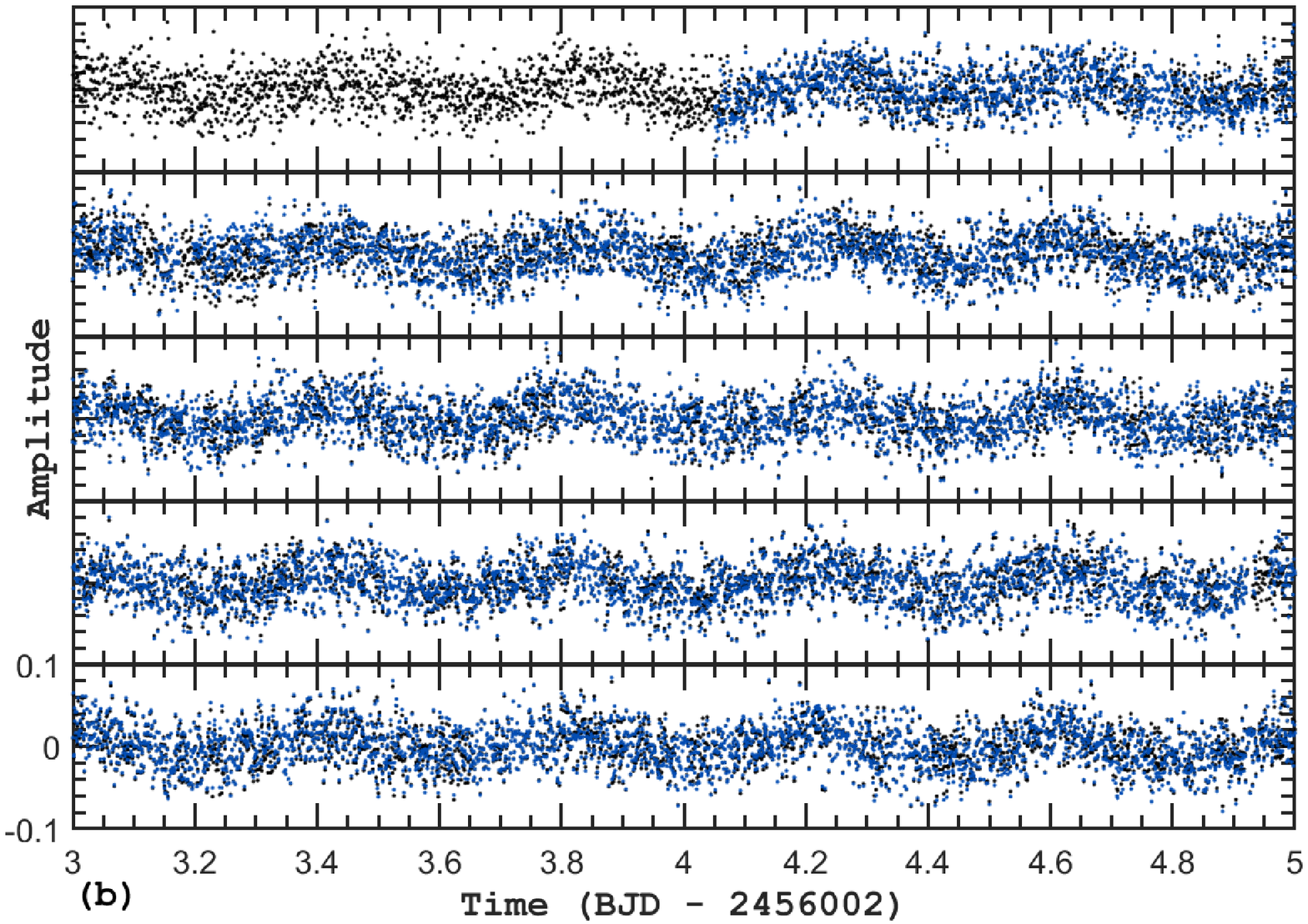}
\includegraphics[width=8.5cm]{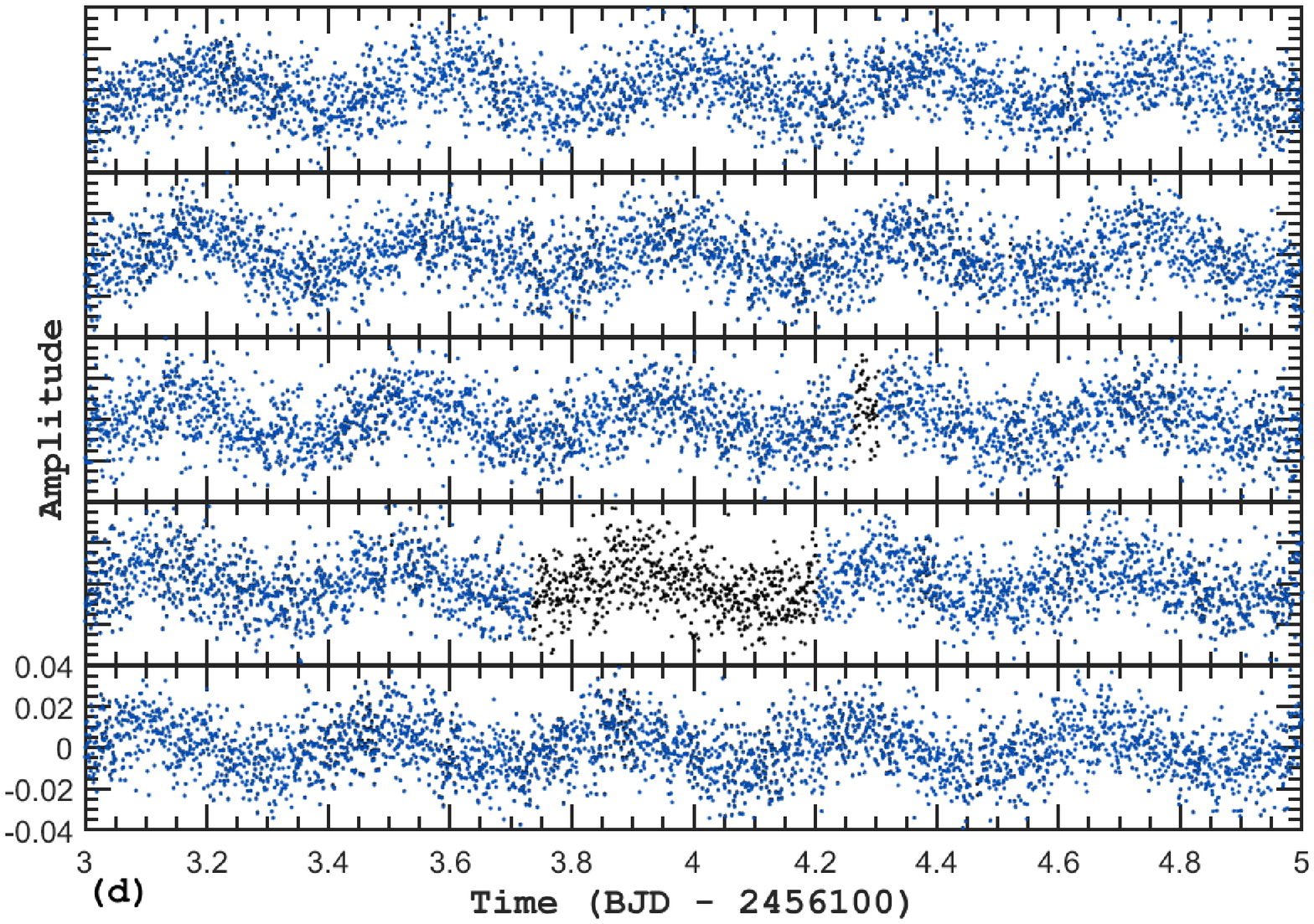}
\caption{Assembled light curves (amplitude is the residual flux relative to the mean brightness of the star vs. time in relative Barycentric Julian Date) of KIC\,2438324 (a) and KIC\,11179657 (c) from the {\sl Kepler} observations. Expanded views show ten days of the light curves by slices of two days for KIC\,2438324 (b) and KIC\,11179657 (d). Raw and corrected fluxes are in black and dark blue, respectively. These colors are also adopted for the following figures\,\ref{flsp} to \ref{doublet2}.
\label{flc}}
\end{figure*}

\begin{figure*}
\centering
\includegraphics[width=8.5cm]{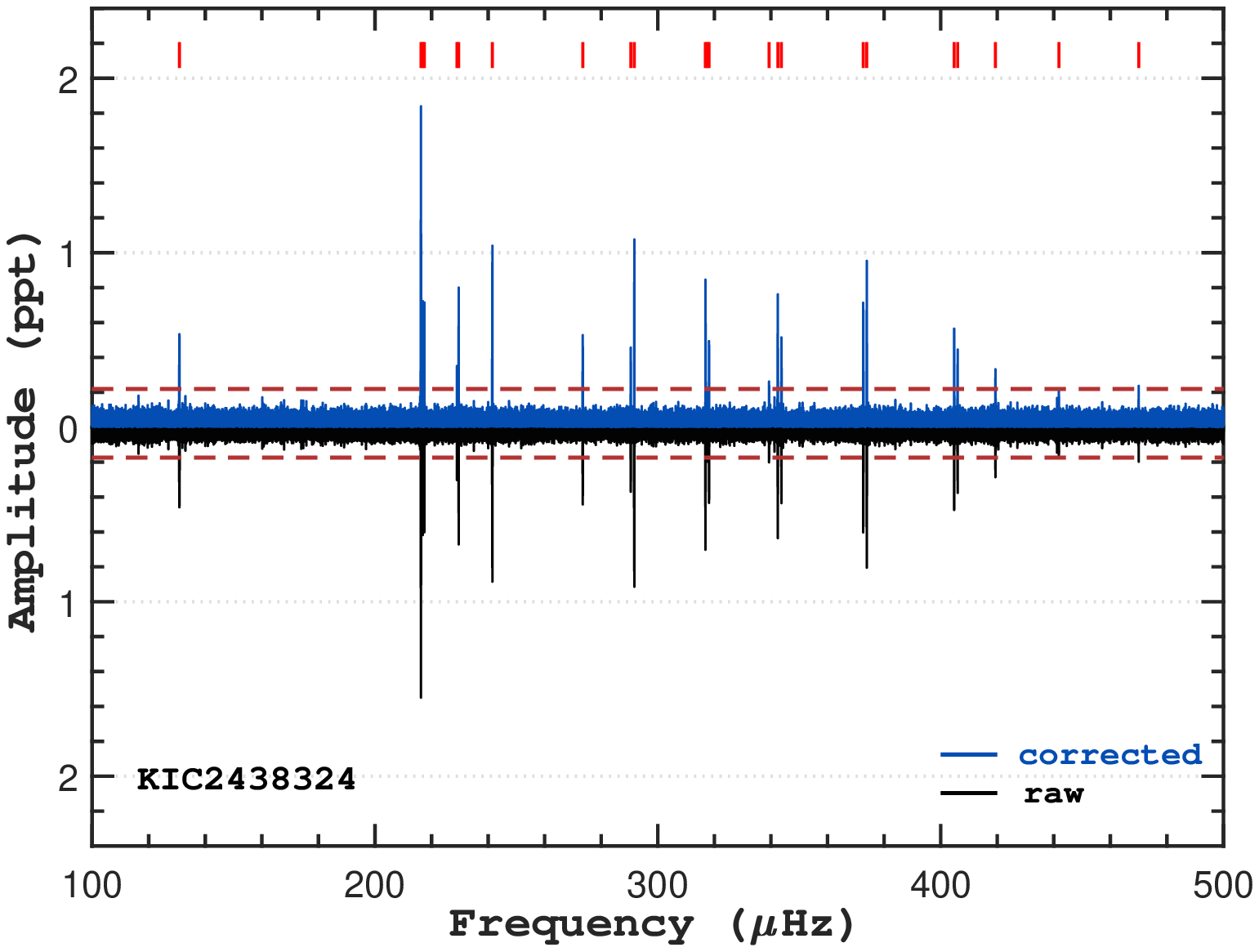}
\includegraphics[width=8.5cm]{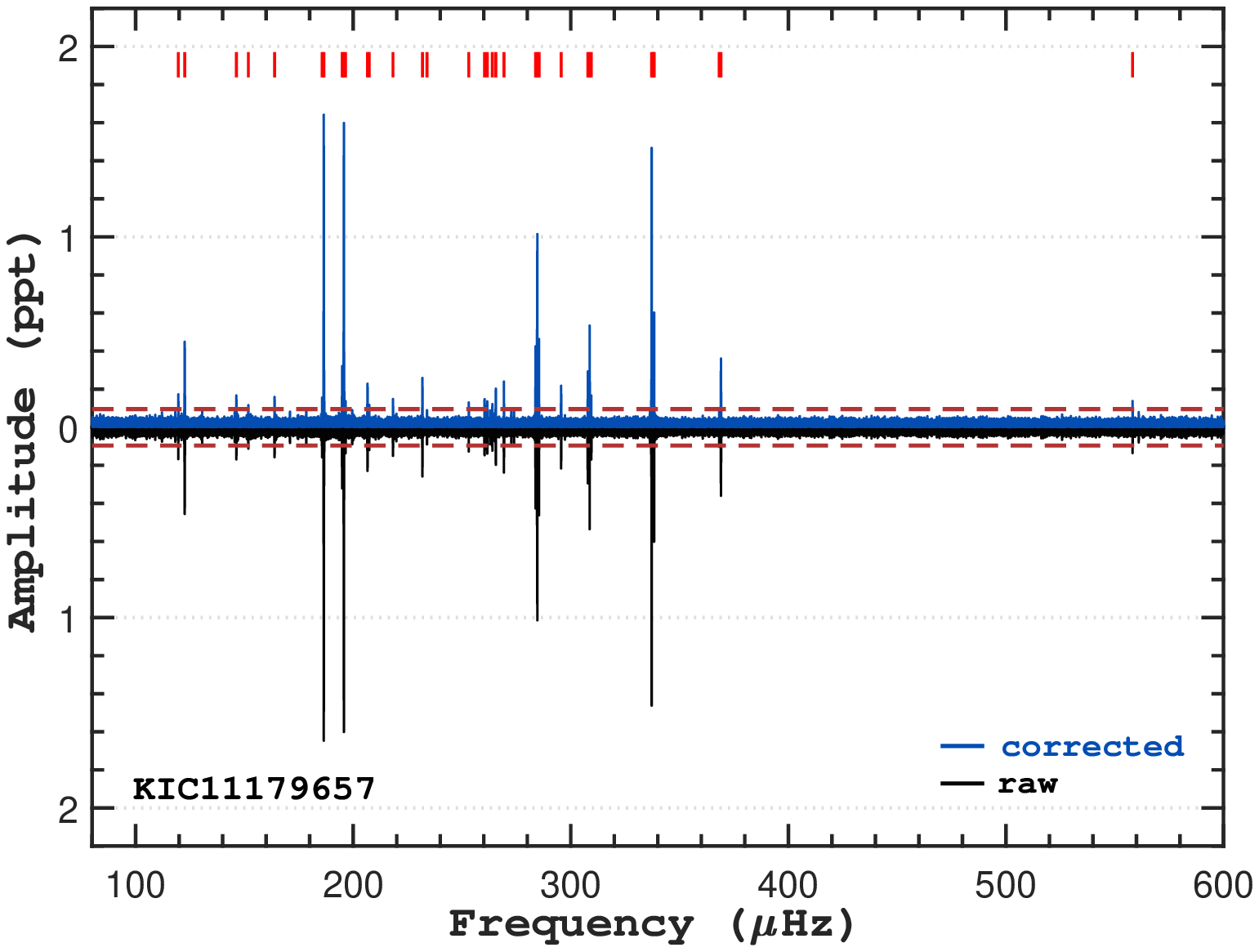}
\caption{Parts of LSP of the {\sl Kepler} photometry collected on KIC\,2438324 ({\sl right}) and KIC\,11179657 ({\sl left}), covering the frequency range where the pulsation signals are found. The vertical segments locate the position of the detected signals. The dashed horizontal lines are the 5.6$\sigma$ threshold of the local noise. A distinct feature of amplitude difference is clearly seen between the raw (upside-down) and corrected LSP of KIC\,2438324 but not KIC\,11179657.
\label{flsp}}
\end{figure*}

In the following, both raw and corrected light curves are analyzed. These were produced through the standard Kepler Science Processing Pipeline provided by \citet{2010ApJ...713L..87J}. As {in} Paper\,\RNum{1}, we performed additional data detrending and $3\sigma$-clipping to remove residual drifts and data point outliers. After these operations, the raw and corrected light curves for KIC\,11179657 consist of 1,413,308 and 1,399,858 points, with duty cycles of 91.6\% and 90.7\%, {respectively,} over a period of $\sim2.88$ years.
For KIC\,2438324, the light curves contain 1,227,891 and 1,235,149 measurements for the corrected and raw flux, corresponding to duty cycles of 72.9\% and 73.3\%, {respectively,} over $\sim3.14$\,years. The two light curves are shown {in their entirety on the top} panels of Figure\,\ref{flc}. A clear feature related to KIC\,2438324 is that light variations in the corrected flux are significantly higher than that of {the} raw flux when the contamination factor is large (see Table\,\ref{t1}). Parts of the light curves are expanded in bottom panels where the brightness variations from the binary reflection effect dominate over the pulsations. The latter are however clearly seen in the Lomb-Scargle Periodograms \citep[LSP;][]{lomb76,scargle82} represented in Figure\,\ref{flsp}. The two stars show very similar LSPs with oscillations mostly in the low-frequency {\sl g}-mode domain between 100 and 400\,$\mu$Hz, and few low-amplitude modes in the 400 -- 600\,$\mu$Hz range. However, the noise level for KIC\,11179657 is much lower than that of KIC\,2438324, although their main frequencies have very similar amplitudes, $\sim2$\,ppt. 
We note that the amplitudes of oscillations in KIC\,2438324 differ in the two types of fluxes, mainly due to contaminating light from the nearby star.

\startlongtable
\begin{deluxetable*}{crcrccrrrc} \label{sdbt2}
\centering
\tablecaption{List of frequencies detected in KIC~2438324. \label{t2}}
\tablehead{
\colhead{Id.} & \colhead{Frequency} & \colhead{$\sigma_\mathrm{f}$} & \colhead{Period} & \colhead{$\sigma_\mathrm{P}$}
& \colhead{Amplitude} & \colhead{$\sigma_\mathrm{A}$}  & \colhead{S/N} & \colhead{Modulation\tablenotemark{a}}  \\
\colhead{} & \colhead{($\mu$Hz)} & \colhead{($\mu$Hz)} & \colhead{(s)} & \colhead{(s)}
& \colhead{(ppt)} & \colhead{(ppt)}  & \colhead{} & \colhead{}  }
\startdata
$f_\mathrm{orb}$   &   29.044187  & 0.000018   &  34430.297397  &   0.021593   &   13.383   & 0.0040   & 333.1   &       AM  \\
$2f_\mathrm{orb}$  &   58.088335  & 0.000113   &  17215.160472  &   0.033617   &    2.149   & 0.0040   &  53.5   &       AM  \\
$f_{12}$   &   130.864047    &   0.000442   &     7641.518249   &   0.025805   &    0.545   &  0.040   &  13.7   &       AM  \\
$f_{01}$   &   216.283646    &   0.000129   &     4623.558085   &   0.002761   &    1.837   &  0.039   &  47.0   &       AFM  \\
$f_{10}$   &   216.891302    &   0.000346   &     4610.604431   &   0.007357   &    0.685   &  0.039   &  17.5   &       AFM  \\
$f_{09}$   &   217.487634    &   0.000342   &     4597.962562   &   0.007240   &    0.692   &  0.039   &  17.7   &       AFM  \\
$f_{18}$   &   229.002081    &   0.000656   &     4366.772541   &   0.012516   &    0.360   &  0.039   &   9.2   &       ...  \\
$f_{06}$   &   229.606249    &   0.000296   &     4355.282166   &   0.005623   &    0.796   &  0.039   &  20.5   &       AM  \\
$f_{03}$   &   241.516688    &   0.000223   &     4140.500628   &   0.003821   &    1.058   &  0.039   &  27.2   &       AM  \\
$f_{14}$   &   273.479046    &   0.000451   &     3656.587279   &   0.006035   &    0.519   &  0.039   &  13.4   &       AM  \\
$f_{16}$   &   290.442213    &   0.000520   &     3443.025692   &   0.006159   &    0.450   &  0.039   &  11.7   &       ...  \\
$f_{02}$   &   291.709724    &   0.000216   &     3428.065358   &   0.002535   &    1.083   &  0.039   &  28.1   &       AM  \\
$f_{05}$   &   316.834658    &   0.000275   &     3156.220361   &   0.002738   &    0.848   &  0.038   &  22.1   &       AM  \\
$f_{22}$   &   317.483152    &   0.001047   &     3149.773437   &   0.010388   &    0.223   &  0.038   &   5.8   &       ...  \\
$f_{15}$   &   318.110788    &   0.000456   &     3143.558902   &   0.004508   &    0.511   &  0.038   &  13.3   &       AM  \\
$f_{20}$   &   339.332074    &   0.000881   &     2946.965748   &   0.007654   &    0.264   &  0.038   &   6.9   &       ...  \\
$f_{07}$   &   342.428859    &   0.000304   &     2920.314614   &   0.002597   &    0.762   &  0.038   &  19.9   &       AFM  \\
$f_{13}$   &   343.691933    &   0.000443   &     2909.582399   &   0.003751   &    0.524   &  0.038   &  13.7   &       AFM  \\
$f_{08}$   &   372.630047    &   0.000318   &     2683.626853   &   0.002290   &    0.730   &  0.038   &  19.1   &       AFM  \\
$f_{04}$   &   373.893293    &   0.000243   &     2674.559880   &   0.001735   &    0.957   &  0.038   &  25.0   &       AFM  \\
$f_{11}$   &   404.773515    &   0.000412   &     2470.517370   &   0.002513   &    0.563   &  0.038   &  14.7   &       AM  \\
$f_{17}$   &   406.037063    &   0.000515   &     2462.829359   &   0.003122   &    0.450   &  0.038   &  11.8   &       AM  \\
$f_{19}$   &   419.376109    &   0.000685   &     2384.494441   &   0.003896   &    0.337   &  0.038   &   8.9   &       ...  \\
$f_{21}$   &   470.044374    &   0.000960   &     2127.458713   &   0.004344   &    0.240   &  0.038   &   6.3   &       ...  \\
\enddata
\tablenotetext{a}{Note: Amplitude (AM) and amplitude/frequency (AFM) modulation provided for that frequency. See text for details.}

\end{deluxetable*}

\startlongtable
\begin{deluxetable*}{crcrcccrcc} 
\centering
\tablecaption{List of frequencies detected in KIC~1117657. \label{t3}}
\tablehead{
\colhead{Id.} & \colhead{Frequency} & \colhead{$\sigma_\mathrm{f}$} & \colhead{Period} & \colhead{$\sigma_\mathrm{P}$}
& \colhead{Amplitude} & \colhead{$\sigma_\mathrm{A}$}  & \colhead{S/N} & \colhead{Modulation\tablenotemark{a}}  \\
\colhead{} & \colhead{($\mu$Hz)} & \colhead{($\mu$Hz)} & \colhead{(s)} & \colhead{(s)}
& \colhead{(ppt)} & \colhead{(ppt)}  & \colhead{} & \colhead{}  }
\startdata     
$f_\mathrm{orb}$   &   29.341850  & 0.000011   &  34081.013757  &   0.012466   &    9.211   & 0.0018   & 518.1   &       AM  \\
$2f_\mathrm{orb}$  &   58.683671  & 0.000149   &  17040.515462  &   0.043206   &    0.661   & 0.0018   &  37.4   &       AM  \\
$f_{20}$   &   119.619242    &   0.000582   &     8359.859033   &   0.040645   &    0.168   &  0.018   &   9.6   &       ...  \\
$f_{07}$   &   122.555274    &   0.000212   &     8159.583591   &   0.014084   &    0.463   &  0.018   &  26.3   &       ...  \\
$f_{21}$   &   146.298704    &   0.000588   &     6835.330534   &   0.027491   &    0.166   &  0.018   &   9.5   &       ...  \\
$f_{32}$   &   151.816132    &   0.000857   &     6586.915297   &   0.037182   &    0.114   &  0.018   &   6.5   &       ...  \\
$f_{22}$   &   163.867034    &   0.000618   &     6102.508697   &   0.023010   &    0.157   &  0.017   &   9.0   &       ...  \\
$f_{25}$   &   185.725763    &   0.000662   &     5384.282638   &   0.019193   &    0.147   &  0.017   &   8.4   &       ...  \\
$f_{01}$   &   186.484826    &   0.000058   &     5362.366591   &   0.001682   &    1.659   &  0.017   &  95.1   &       AM  \\
$f_{11}$   &   194.925628    &   0.000291   &     5130.161738   &   0.007651   &    0.333   &  0.017   &  19.1   &       AFM  \\
$f_{02}$   &   195.724900    &   0.000060   &     5109.211964   &   0.001574   &    1.608   &  0.017   &  92.2   &       AFM  \\
$f_{30}$   &   196.505979    &   0.000765   &     5088.903694   &   0.019817   &    0.127   &  0.017   &   7.3   &       ...  \\
$f_{15}$   &   206.580304    &   0.000424   &     4840.732551   &   0.009943   &    0.228   &  0.017   &  13.1   &       ...  \\
$f_{31}$   &   207.379992    &   0.000826   &     4822.065951   &   0.019216   &    0.117   &  0.017   &   6.7   &       ...  \\
$f_{24}$   &   218.282801    &   0.000657   &     4581.212980   &   0.013785   &    0.147   &  0.017   &   8.5   &       ...  \\
$f_{13}$   &   231.820585    &   0.000373   &     4313.680766   &   0.006944   &    0.258   &  0.017   &  14.9   &       ...  \\
$f_{34}$   &   233.913634    &   0.001002   &     4275.082137   &   0.018305   &    0.096   &  0.017   &   5.6   &       ...  \\
$f_{29}$   &   253.113331    &   0.000738   &     3950.799422   &   0.011517   &    0.130   &  0.017   &   7.5   &       ...  \\
$f_{23}$   &   260.392034    &   0.000644   &     3840.363259   &   0.009503   &    0.149   &  0.017   &   8.6   &       ...  \\
$f_{26}$   &   261.632760    &   0.000704   &     3822.151326   &   0.010288   &    0.136   &  0.017   &   7.9   &       ...  \\
$f_{28}$   &   263.969777    &   0.000730   &     3788.312483   &   0.010480   &    0.132   &  0.017   &   7.6   &       ...  \\
$f_{18}$   &   265.556190    &   0.000473   &     3765.681381   &   0.006706   &    0.203   &  0.017   &  11.8   &       ...  \\
$f_{14}$   &   269.269969    &   0.000417   &     3713.744995   &   0.005757   &    0.230   &  0.017   &  13.3   &       ...  \\
$f_{09}$   &   283.831342    &   0.000226   &     3523.219084   &   0.002802   &    0.426   &  0.017   &  24.6   &       AFM  \\
$f_{04}$   &   284.629468    &   0.000095   &     3513.339662   &   0.001169   &    1.014   &  0.017   &  58.7   &       AFM  \\
$f_{08}$   &   285.409755    &   0.000210   &     3503.734486   &   0.002581   &    0.457   &  0.017   &  26.4   &       AFM  \\
$f_{16}$   &   295.574098    &   0.000434   &     3383.246395   &   0.004964   &    0.221   &  0.017   &  12.8   &       ...  \\
$f_{12}$   &   307.855429    &   0.000330   &     3248.277947   &   0.003484   &    0.290   &  0.017   &  16.8   &       AFM  \\
$f_{06}$   &   308.657235    &   0.000175   &     3239.839820   &   0.001838   &    0.547   &  0.017   &  31.8   &       AFM  \\
$f_{19}$   &   309.438786    &   0.000549   &     3231.656942   &   0.005736   &    0.174   &  0.017   &  10.1   &       ...  \\
$f_{03}$   &   337.173926    &   0.000065   &     2965.828388   &   0.000575   &    1.456   &  0.017   &  85.0   &       AM  \\
$f_{17}$   &   337.954922    &   0.000441   &     2958.974514   &   0.003863   &    0.216   &  0.017   &  12.6   &       ...  \\
$f_{05}$   &   338.298358    &   0.000162   &     2955.970598   &   0.001418   &    0.587   &  0.017   &  34.3   &       AM  \\
$f_{33}$   &   368.217265    &   0.000878   &     2715.787918   &   0.006474   &    0.108   &  0.017   &   6.3   &       ...  \\
$f_{10}$   &   369.030407    &   0.000263   &     2709.803799   &   0.001930   &    0.361   &  0.017   &  21.2   &       AM  \\
$f_{27}$   &   558.210809    &   0.000692   &     1791.437901   &   0.002220   &    0.136   &  0.017   &   8.0   &       ...  \\
\enddata
\tablenotetext{a}{Note: Amplitude (AM) and amplitude/frequency (AFM) modulation provided for that frequency. See text for details.}

\end{deluxetable*}

Frequency extraction follows the same procedure as in Paper\,\RNum{1}, which is based on a standard prewhitening and nonlinear least-squares fitting approach \citep{deeming75}. We provide in Table\,\ref{t2} and \ref{t3} the lists of detected peaks for KIC\,2438324 and KIC\,11179657 {respectively} along with their fitted attributes : frequency (in $\mu$Hz; and period in seconds), amplitude (in ppt, i.e., parts per thousand) relative to the mean brightness of the star, and signal-to-noise ratio (S/N) of the detection. Each attribute has an associated error : $\sigma_\mathrm{f}$, $\sigma_\mathrm{p}$, and $\sigma_\mathrm{A}$. The ``ID'' column uniquely identifies a detected frequency with a sequence number that indicates its rank by order of decreasing amplitude. The ``Modulation'' column indicates if amplitude only (AM: {for characterization of systematic modulations}) or amplitude/frequency modulations (AFM: {in representative rotational multiplets}) are presented here for a given frequency. These modes are further discussed in the following sections.

Table\,\ref{t2} lists 22 detected frequencies in KIC\,2438324 associated to stellar oscillations and two additional low-frequency peaks of orbital nature. All of them are above an adopted threshold of 5.6 $\times$ the noise level \citep[5.6$\sigma$ as tested in][]{zong2016a}. {There are} 16 frequencies {attributed} to components of two triplets near 217 and 317\,$\mu$Hz, and five doublets near 229, 291, 342, 372 and 406\,$\mu$Hz. One component, $f_{16}$, of the doublet near 290\,$\mu$Hz falls close to ten times the orbit frequency, $f_\mathrm{orb}\sim29~\mu$Hz. We note that all frequencies are found in the 100--500\,$\mu$Hz range. 
For KIC\,11179657 (Table\,\ref{t3}), we detected 33 independent frequencies above 5.6$\sigma$, all attributed to oscillations. We identify three triplets near 195, 284 and 307\,$\mu$Hz, four doublets near 185, 206, 337 and 368\,$\mu$Hz, and a possible quintuplet near 260\,$\mu$Hz with two missing components. Table\,\ref{t3} also contains a frequency, $f_{32}$, which is close to the difference of $f_{05}$ and $f_{01}$. We further note that two weak peaks (of $S/N\sim5$; i.e., not provided in Table\,\ref{t3}) are seen at 336.37\,$\mu$Hz and 262.88\,$\mu$Hz. These could be additional components belonging to the doublet near 337\,$\mu$Hz and the 260\,$\mu$Hz quintuplet, respectively. In addition to the oscillations, we also detected a binary signal, $f_\mathrm{orb}$, at a frequency of 29.341850\,$\mu$Hz and its first harmonic 2$f_\mathrm{orb}$. Like for KIC\,2438324, most of the frequencies of KIC\,11179657 are found in the 100--400\,$\mu$Hz frequency range, except the frequency $f_{27}$ at 558.2~$\mu$Hz.

\section{Modulation patterns}

Following the same strategy as in Paper\,\RNum{1} to analyze amplitude and frequency modulations in these two stars, we focus on frequencies of highest amplitudes, in particular those being components of rotational multiplets. We note that, for both stars, there are relatively few frequencies compared to KIC\,3527751 discussed in Paper\,\RNum{1}. Contrary to paper\,\RNum{1}, we do not provide sliding LS-Periodograms (sLSP) due to {two specific limitations}: (1) In general, we find that most frequencies are close to stability and do not show obvious variations in these diagrams, {as they} are not sensitive enough. (2) They are of limited use to precisely measure modulation patterns for comparison between raw and corrected fluxes (the main objective of the present analysis). In the following, we therefore only present the modulations obtained by prewhitening the frequencies in various parts of the light curve. The 35-month light curve of KIC\,2438324 was divided into a series of adjacent pieces using a time step of about 10~d and a filtering window of width 180~d. Each part was analyzed using the same prewhitening technique applied to the whole light curve. This provided 95 measurements for all frequencies with amplitudes above $\sim0.7$~ppt. More details on how to obtain measurements from each light-curve segment can be found in Section~4.2 of Paper\,\RNum{1}. For KIC\,11179657, the light curve was divided into pieces of time step $\sim15$~d and window width of 120~d. The difference of time step (10~d vs 15~d) does not change the general patterns of the amplitude and frequency modulations, but using the longer time step was necessary to reduce computation time. The time window size of 120~d was chosen to ensure sufficient resolution in frequency and to minimize contamination from side-band frequencies associated to the orbital signal with very large amplitude that suffers instrumental amplitude modulations. Since this star experienced two long interruptions of its monitoring, a few measurements were discarded when they were very close to large interval gaps. We finally obtained 45 measurements for the frequencies of amplitude above $\sim0.3$~ppt. 

\subsection{Representative frequencies in KIC\,2438324}
\begin{figure*}
\centering
\includegraphics[width=2.89cm]{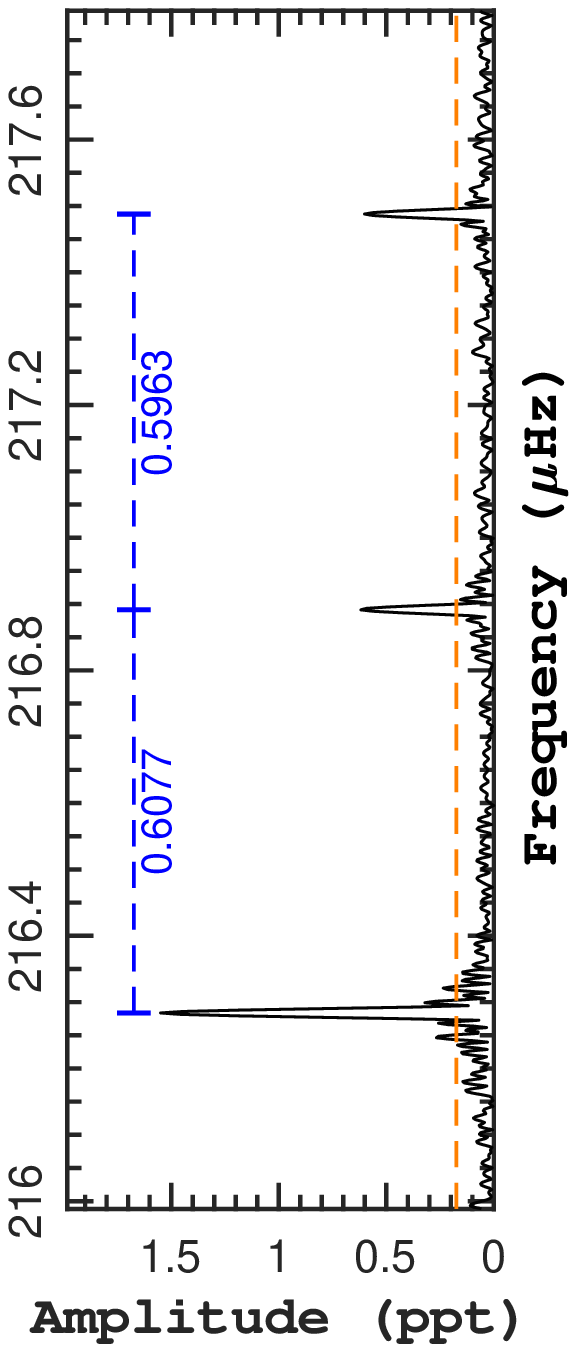}\includegraphics[width=12cm]{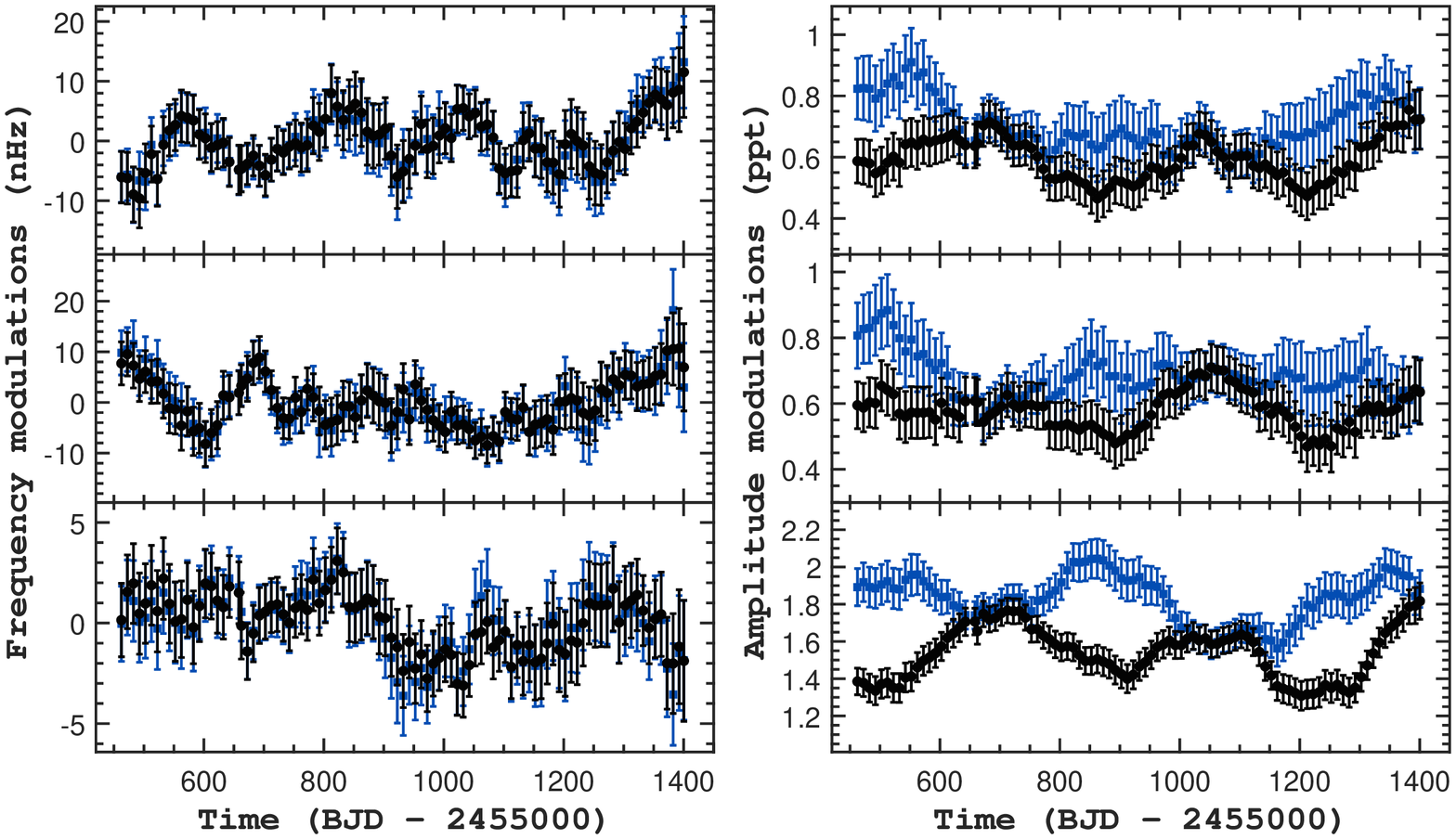}
\caption{Amplitude (AM) and frequency modulations (FM) of the triplet components near 217\,$\mu$Hz in KIC\,2438324. {\sl Left} panel: the LSP shows nearly equidistant structures whose frequency spacing values are given in the text. The dashed line is 
our typical detection limit of 5.6$\sigma$ above local median noise. The corresponding FMs and AMs of each components, measured from two different correction levels of the photometric data, are presented in the {\sl middle} and {\sl right} panels, respectively. Note that the FMs are shifted to their averaged values.
\label{triplet}}
\end{figure*}

In the following, we do not provide AMs and FMs for frequencies with {\sl S/N} lower than $\sim12$, due to their small number of exploitable measurements. We end up with 16 frequencies that contain more than 50 measurements each. In this section, we focus on AMs and FMs occurring in one triplet and two doublets as representative modes to illustrate the differences encountered with raw and corrected photometric data. For the other modes, the uncovered patterns will be discussed in Section~\ref{sec:c}.

Figure\,\ref{triplet} shows the AMs and FMs detected in the triplet components near 216.9\,$\mu$Hz. The LSP of the full light curve shows rather simple peak structures suggesting that the three components experience only weak modulations. This is confirmed by the precise measurements of the modulation patterns presented in the middle and right panels. The retrograde ($m=-1$) component of the triplet displays frequency variations within $\sim \pm2$\,nano~Hz, i.e., comparable to the uncertainties. The other two components show larger FMs of $\sim\pm10$\,nano~Hz. Besides, during most of the observing run, the FMs appear to be somewhat antiphased between the $m=0$ and $m=-1$ components. Comparing now the two data reduction levels (raw vs corrected flux), we find that FMs of the three components are very similar. However, {the two different types of} AM patterns show {some} differences. For instance, the retrograde mode evolves in apparent antiphase over most of the observations. Despite these differences, we find that these AMs show very similar modulation patterns in each type of flux, overall, {which is further illustrated in Figure\,\ref{sdb2stable}}.

\begin{figure*}
\centering
\includegraphics[width=2.89cm]{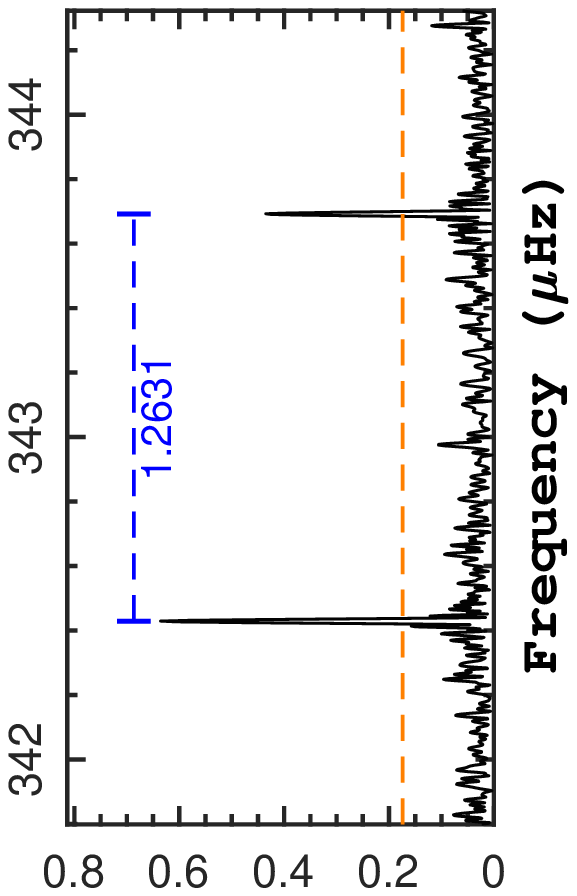}\includegraphics[width=12cm]{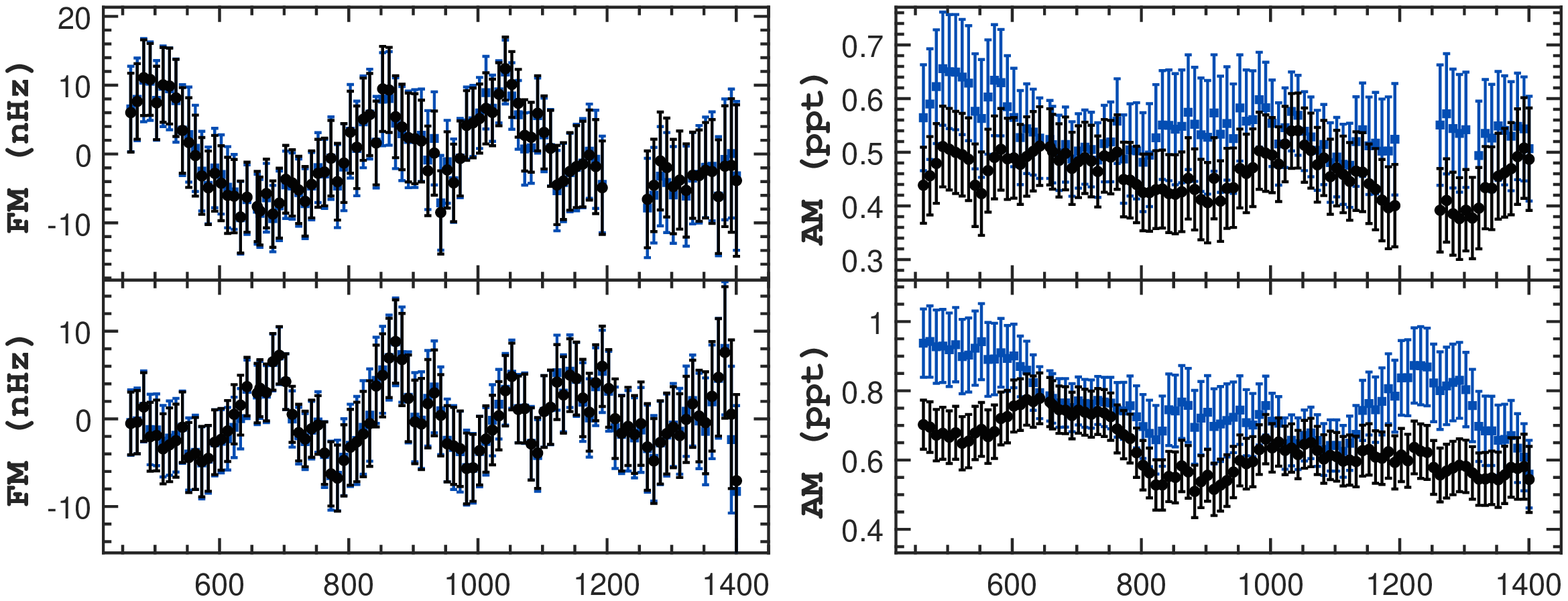}\\
\includegraphics[width=2.89cm]{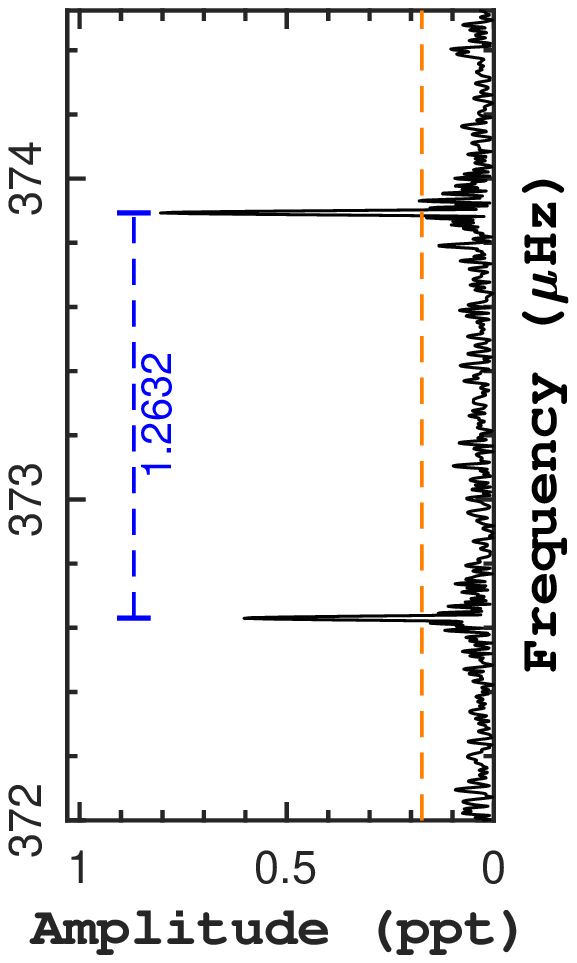}\includegraphics[width=12cm]{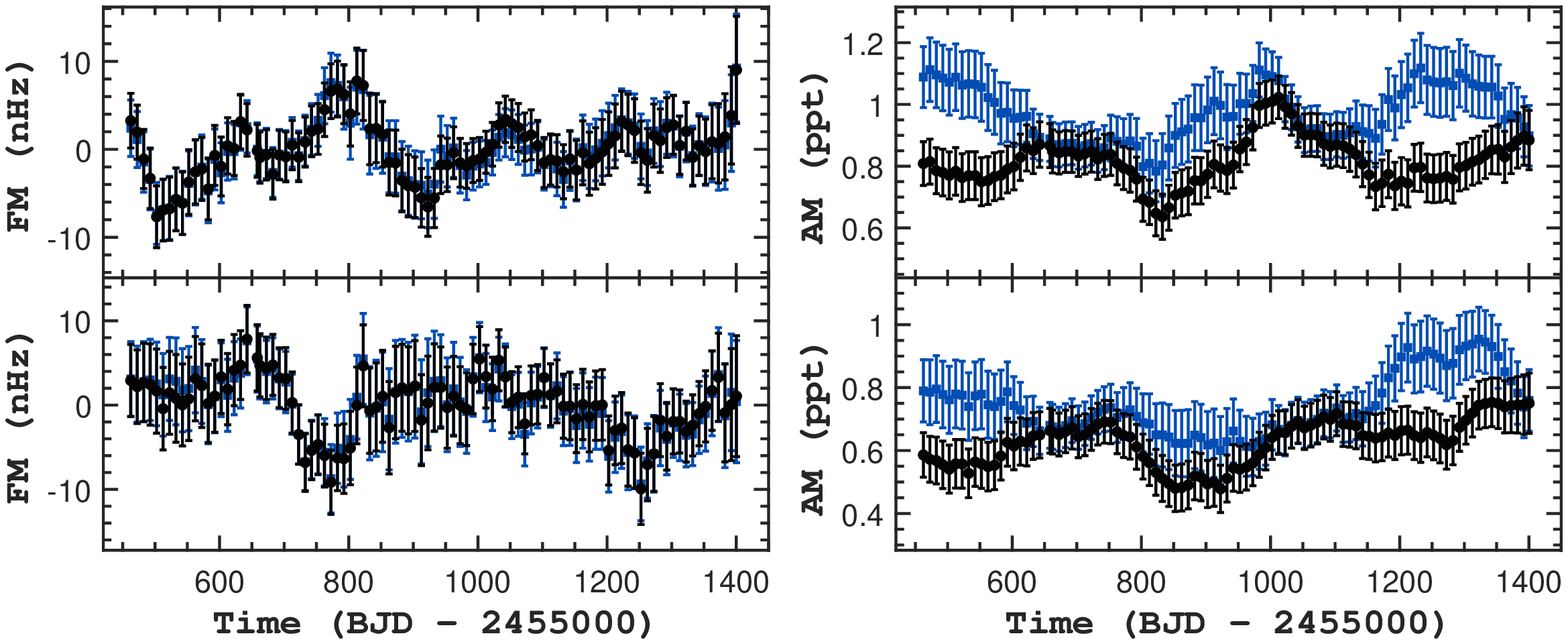}\\
\caption{Same to Figure\,\ref{triplet} but for an example of two doublets near 343 and 373\,$\mu$Hz in KIC\,2438324 (from top to bottom panels), respectively.
\label{doublet}}
\end{figure*}

Figure\,\ref{doublet} illustrates the AMs and FMs disclosed in the components forming two doublets near 343 and 373\,$\mu$Hz, respectively. The structures in the LSP near these two doublets have, again, relatively simple forms, similar to that of the 216.9\,$\mu$Hz triplet. The two ($m=\pm1$) components forming the 343\,$\mu$Hz doublet clearly exhibit some anti-correlation during most of the observing run, with almost the same values extracted from both types of fluxes. For the FMs occurring in the 373\,$\mu$Hz doublet, as measured from the two types of fluxes, the pulse-shape frequency variations only look anti-phased when the ($m=+1$) component reaches its frequency maximum, but with a phase delay of one month. We find that the four frequency variations are within a range of $\pm10$\,nano~Hz. Concerning measured AMs, they appear clearly different in the raw and corrected fluxes, respectively. However, despite these differences, all of them seem to follow similar patterns. We note that the amplitude uncertainty of the 344\,$\mu$Hz component is {relative} large compared to the other three modes. 

\subsection{Representative frequencies in KIC\,11179657}
\begin{figure*}
\centering
\includegraphics[width=2.89cm]{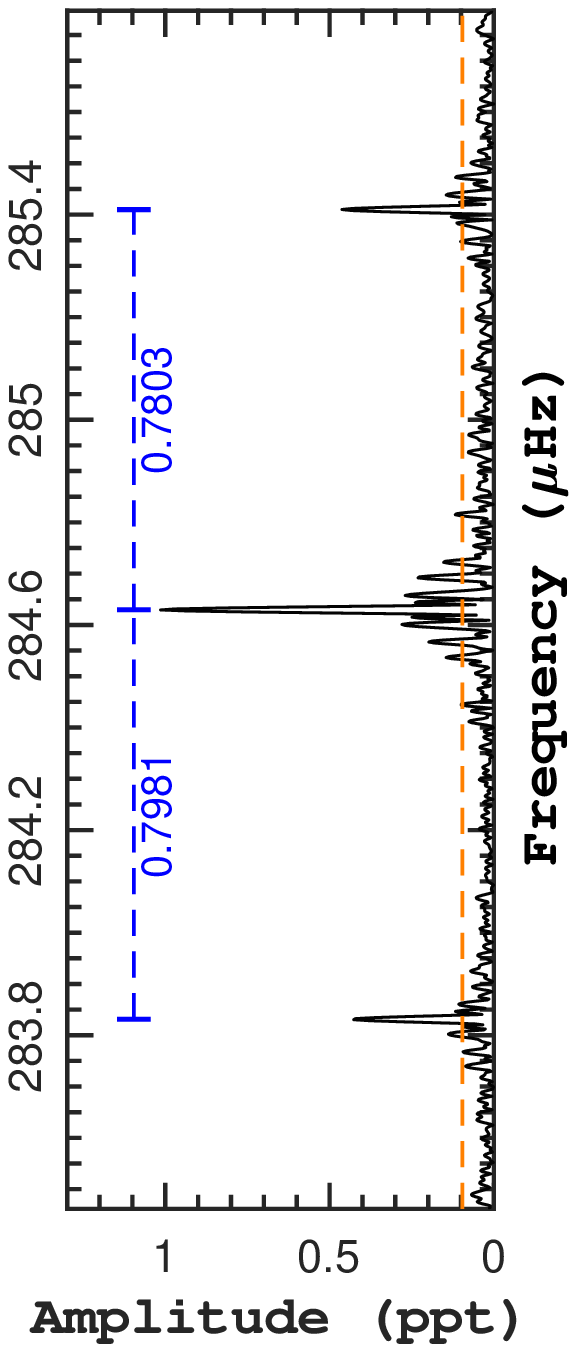}\includegraphics[width=12cm]{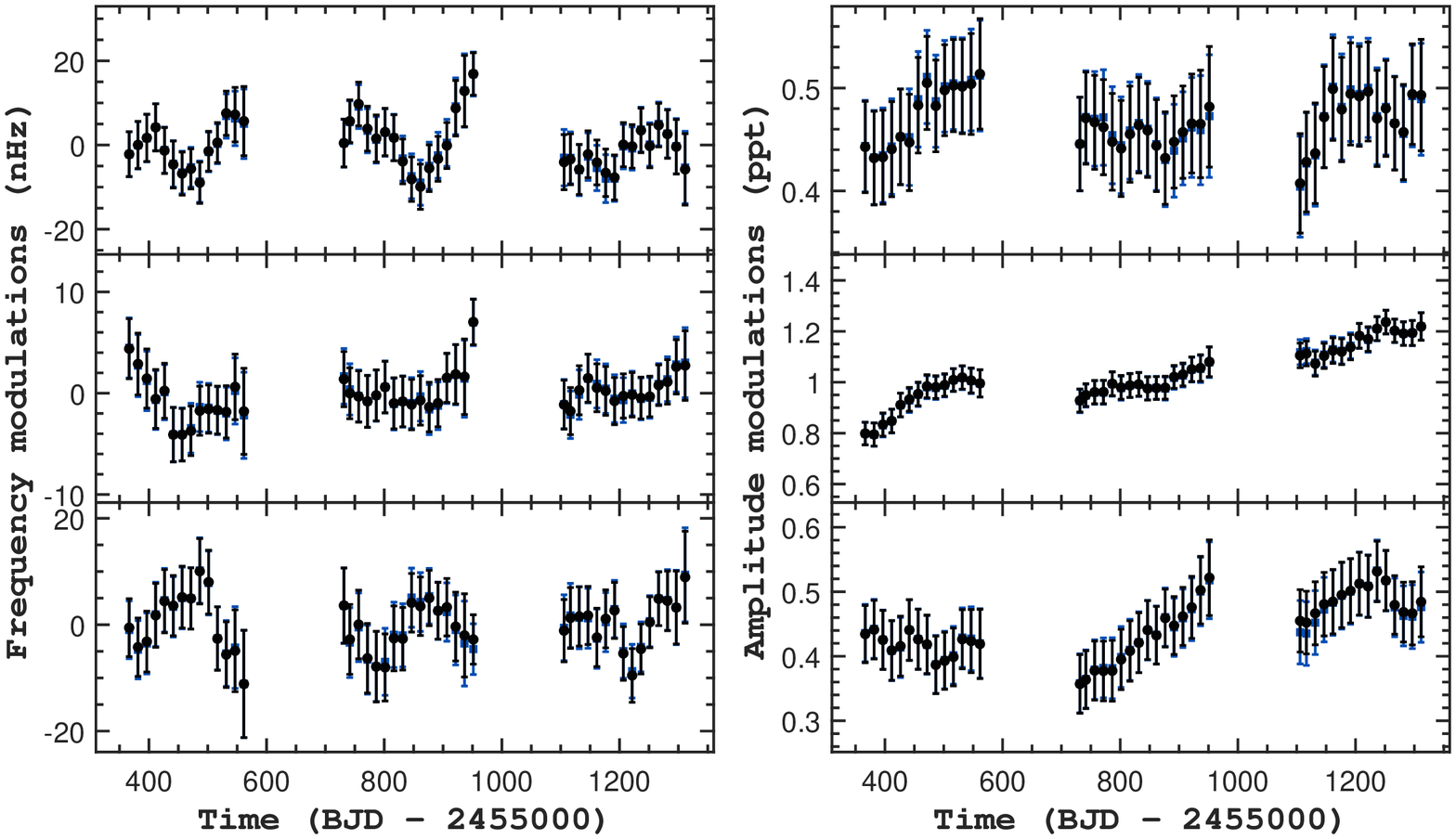}
\caption{Same to Figure\,\ref{triplet} but for an example of triplet near 284.6\,$\mu$Hz in KIC\,11179657. Note that, at this scale, the two different measurements are very close to each others both in amplitude and frequency.
\label{triplet2}}
\end{figure*}

\begin{figure*}
\centering
\includegraphics[width=2.89cm]{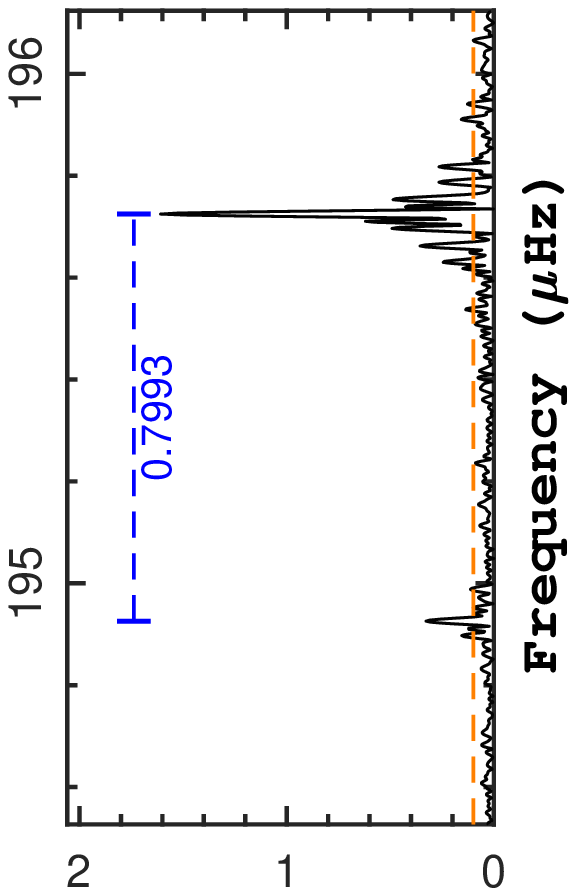}\includegraphics[width=12cm]{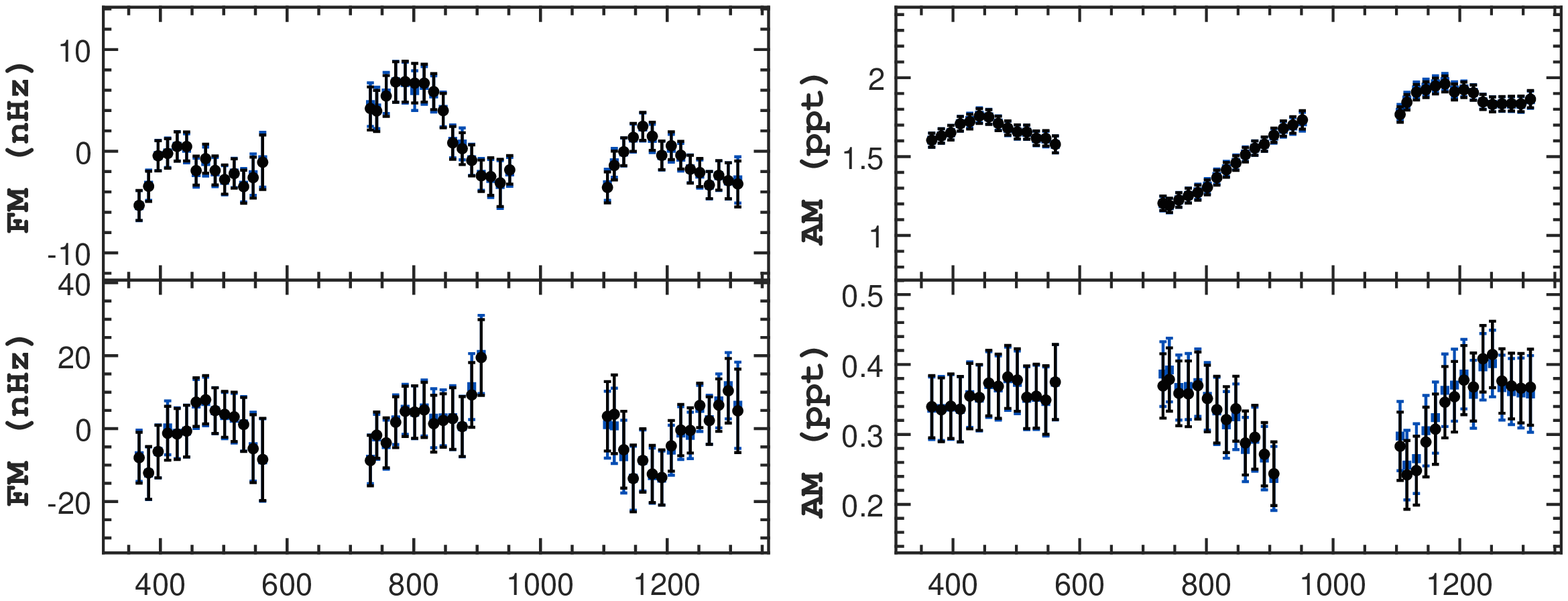}\\
\includegraphics[width=2.89cm]{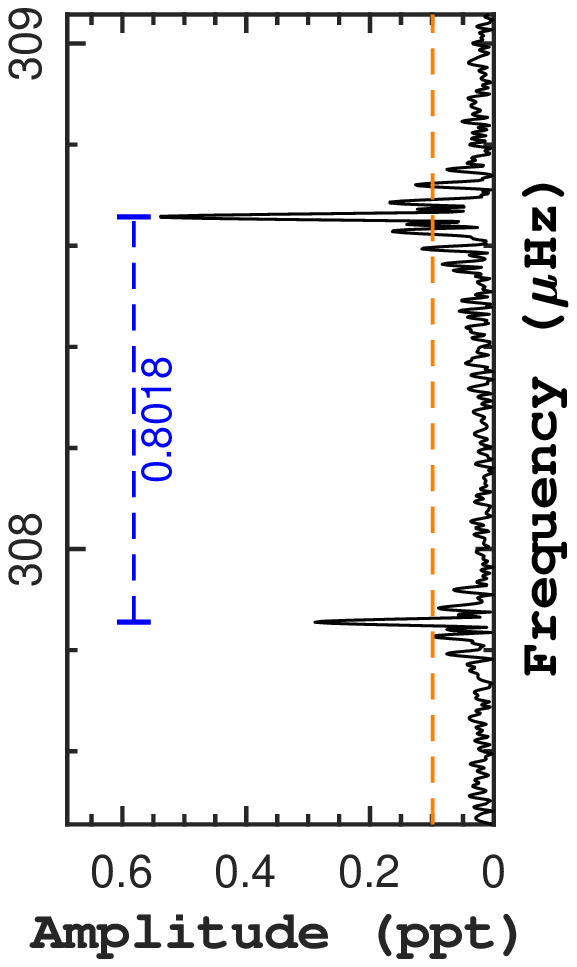}\includegraphics[width=12cm]{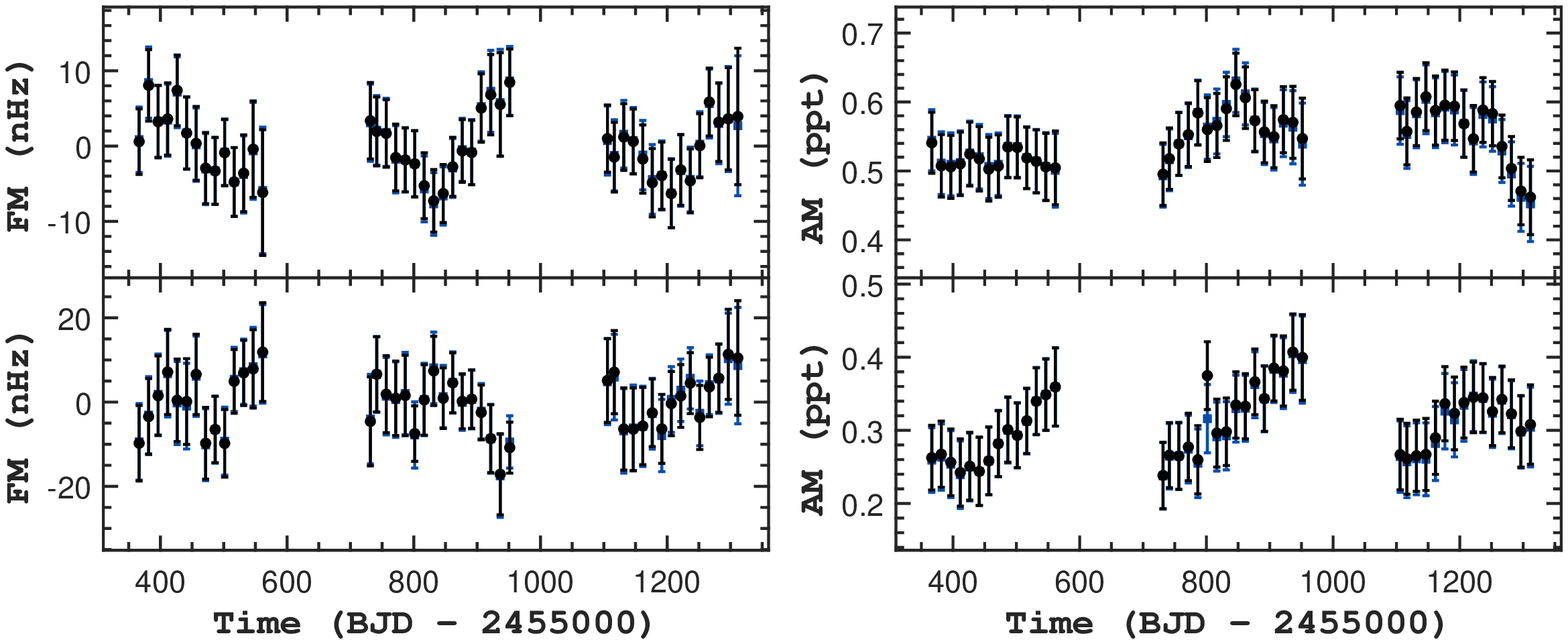}\\
\caption{Same to Figure\,\ref{triplet} but for examples of two triplets near 196 and 308\,$\mu$Hz in KIC\,11179657 (from top to bottom panels), respectively. Note that both triplets have one weak side component whose AM is not provided due to its small amplitude.
\label{doublet2}}
\end{figure*}

Similar to KIC\,2438324, AMs and FMs are provided for one triplet and two doublets as representative frequencies observed in KIC\,11179657. In this case, we measured AMs and FMs for modes with {\sl S/N} values down to about 20, a bit higher than for KIC\,2438324. This establishes 11 modes with measured modulations out of 34 detected frequencies. This higher {\sl S/N} limit is chosen because the light curves run across two large interval gaps. Details on these 11 frequencies with precise AMs and FMs are presented in Section~\ref{sec:c}.

Figure\,\ref{triplet2} shows the AMs and FMs disclosed in the components belonging to the triplet near 284.6\,$\mu$Hz. The nearly equidistant triplet reveals broadened structures in the LSP for each component, due to aliasing introduced by the large gaps in the light curves. A noticeable feature is that both AMs and FMs show almost the same patterns whatever the flux calibration used. The $m=\pm1$ components show regular frequency variations evolving in antiphase for some events. The quasi-periodic FMs can be roughly estimated to be on a timescale of one year, with a magnitude of $\pm10$\,nano~Hz. In contrast, the central ($m=0$) component exhibits a less obvious FM, with a magnitude of $\pm4$\,nano~Hz. The AMs happening in this triplet reveal relatively simple patterns, e.g., the central component following a roughly linear increase in amplitude during the whole observation run. We also note that the $m=-1$ and $m=0$ components have very similar AMs during the last observation segment. However, in the first segment, the similar AMs are found between the $m=+1$ and $m=0$ components.  

Figure\,\ref{doublet2} shows the AMs and FMs obtained for the components forming two doublets near 196 and 308\,$\mu$Hz, respectively. These doublets, similar to the triplet previously described, show no difference in the measurements of amplitude and frequency modulations from the raw and corrected fluxes. The 196\,$\mu$Hz doublet has regular frequency variations, which first evolve in phase (the first segment), then gradually switch to antiphase (the last segment) between the two components. The FMs measured in the 308\,$\mu$Hz doublet show the opposite, by first evolving somewhat in antiphase, then gradually becoming in phase, if we roughly estimate the modulation patterns. All these four components are measured with relatively small FMs, typically within a range of $\pm10$\,nano~Hz. A closer look at the AMs of these components suggests possibly regular modulation patterns. The 196\,$\mu$Hz doublet possibly exhibits anti-correlation between its two components. However, the 308\,$\mu$Hz doublet does not show clear correlations in the variations occurring between the two components.

\section{Comparison with an orbital signal} 
\label{sec:c}
\begin{figure*}
\centering
\includegraphics[width=16cm]{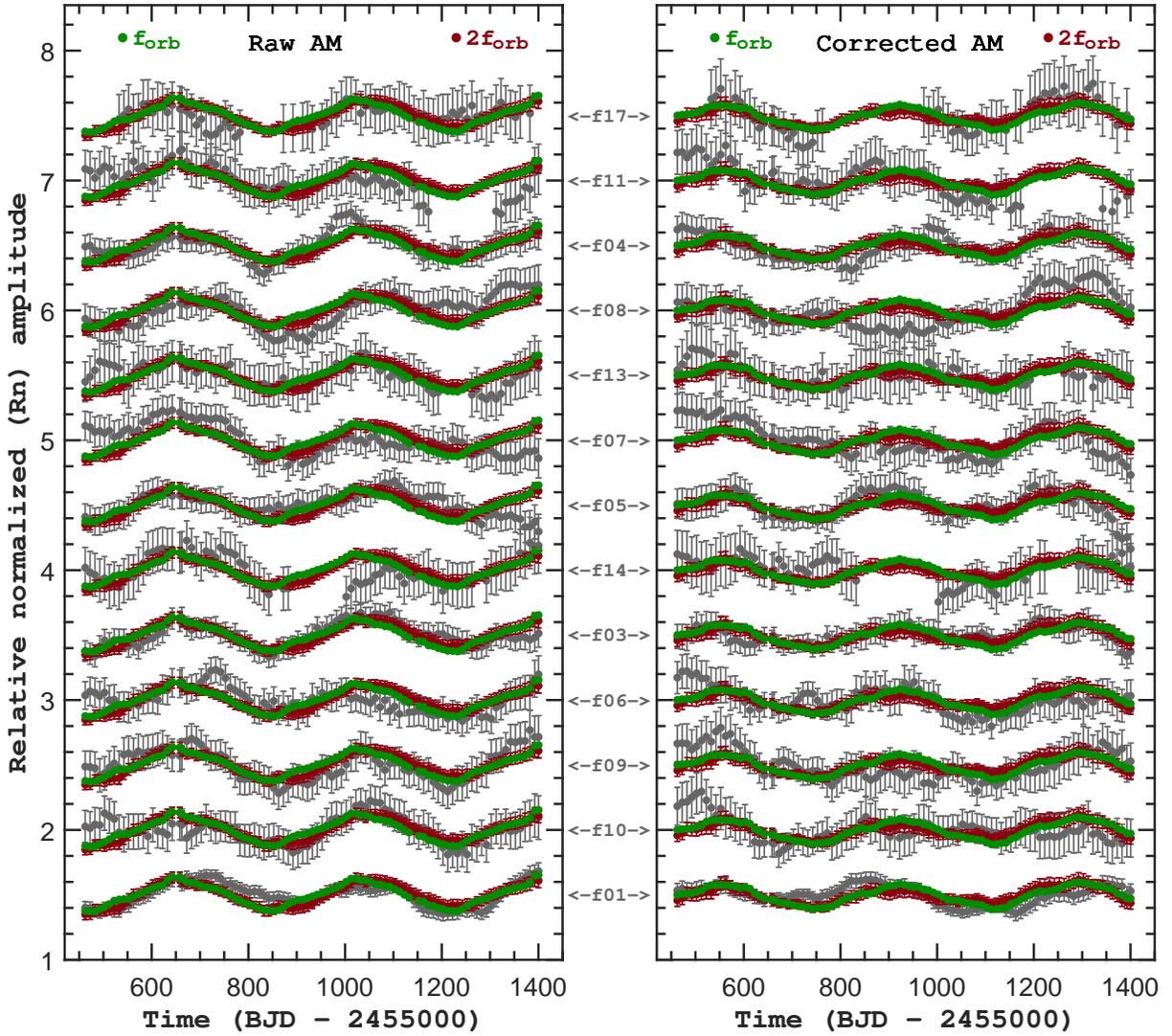}
\caption{AM comparison of 13 independent frequencies with the orbital frequencies in the sdB star KIC\,2438324. {\sl Left panel} shows AMs measured from the raw flux and {\sl right} from the corrected flux. Each of these frequencies are normalized by their averaged amplitudes and shifted to the values where the legends indicate. Note that the errors 
for the orbital frequency $f_\mathrm{orb}$ are smaller than the symbol itself.
\label{sdb2stable}}
\end{figure*}

\begin{figure*}
\centering
\includegraphics[width=16cm]{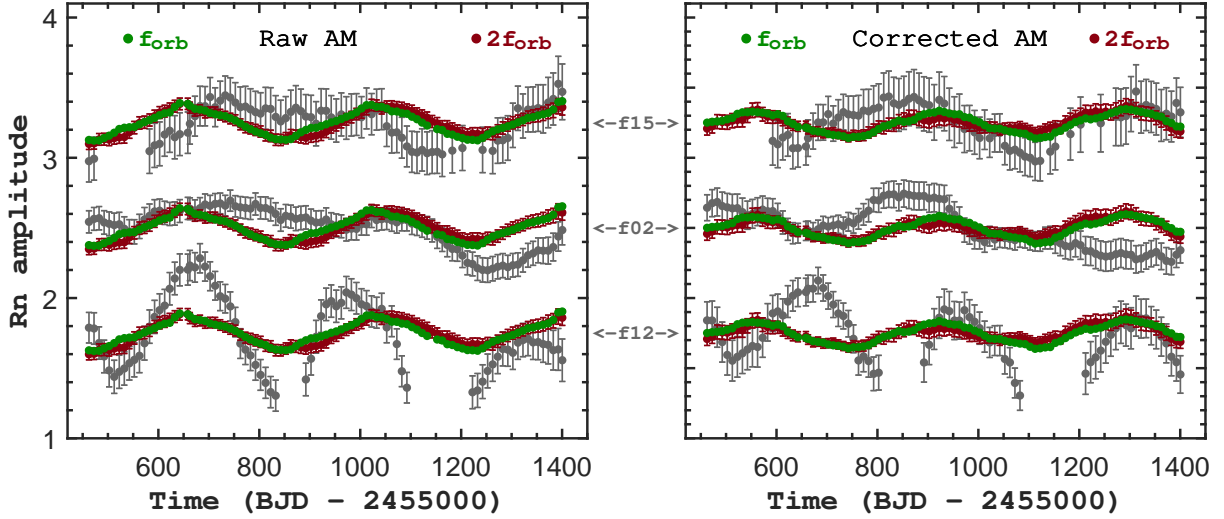}
\caption{Same to Figure\,\ref{sdb2stable} but for another three independent frequencies in KIC\,2438324. {Note that these AMs are significantly different to those in Figure\,\ref{sdb2stable}}
\label{campm}}
\end{figure*}

%++++++++++++++++++++++++++++++++++++++++++++++++++++++++++++++++++++++++
\begin{figure}
\centering
\includegraphics[width=7.2cm]{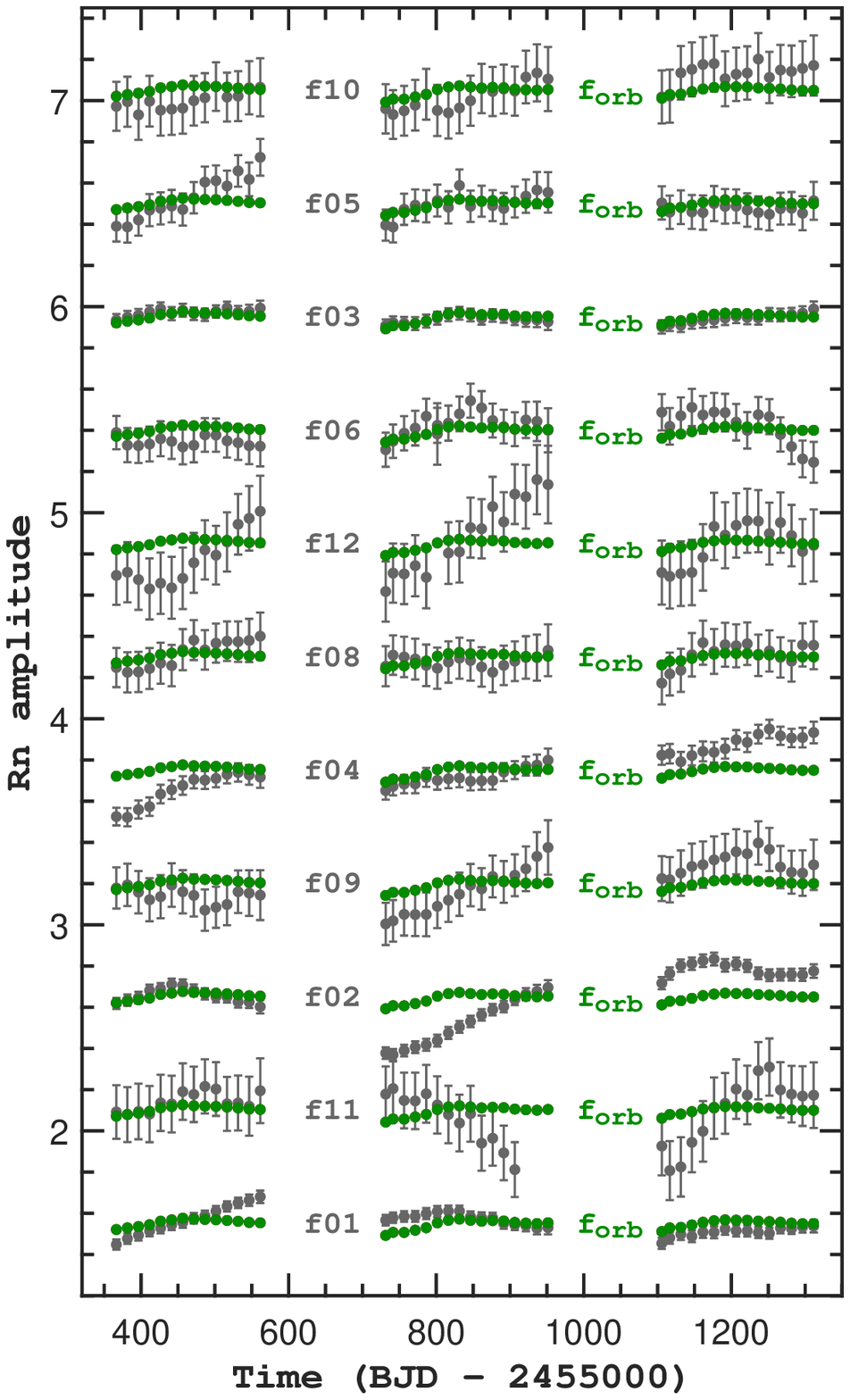}
\caption{AM comparison of 11 independent frequencies with the orbital frequency in the sdB star KIC\,11179657. Each of these frequencies are normalized by their averaged amplitudes and shifted to the values where the legends indicate. Note that the errors 
for the orbital frequency $f_\mathrm{orb}$ are smaller than the symbol itself.
\label{sdb1com}}
\end{figure}
%=========================================================================

%=++++++++++++++++++++++++++++++++++++++++++++++++++++++++++++++++++++++++
%\begin{figure}
%\centering
%\includegraphics[width=7.2cm]{campm1}
%\caption{Similar to Figure\,\ref{sdb1com} but for the comparison of three independent frequencies with the independent frequency $f_{02}$. 
%\label{sdb1com22}}
%\end{figure}
%==========================================================================

As described in the above section, the representative frequencies may have very similar modulation patterns in their amplitudes, which intuitively suggests instrumental effects from the satellite itself. To carefully check for such effects, reliable references should be used as calibrators for the amplitude. Binary signals, such as observed in  KIC\,2438324, can be good candidates for such references, due to their usually high amplitudes (enough precision) and several harmonics (systematic variation) showing up in the LSP. The orbital period is normally stable in a sdB+dM binary system over a time baseline of $\sim3$~yr \citep[see, e.g.,][]{2009AJ....137.3181L}. It is therefore natural to use this binary signal to gauge the variations observed in modes with AMs and FMs.

Figure\,\ref{sdb2stable} shows the comparison of AMs measured in the orbital frequency ($f_{\rm orb}$) and its first harmonic ($2f_{\rm orb}$) with AMs obtained for 13 independent pulsation frequencies of KIC\,2438324. This clearly establishes that $f_\mathrm{orb}$ (and its harmonics) are subject to almost exactly the same modulation patterns, both in the raw and corrected flux data sets. Moreover, in general, the AMs of the independent pulsation frequencies are very similar to that of $f_\mathrm{orb}$, since all measurements are consistent with the AM of $f_\mathrm{orb}$ within the uncertainties. The AMs of $f_\mathrm{orb}$ are measured with peak-to-peak variations of $\sim10~\%$ and $\sim5\,\%$ using the raw and corrected fluxes, respectively. These values also correspond to the magnitude of the variations observed in the 13 independent modes. We also note that the modulation patterns are different, when comparing raw and corrected fluxes, both in terms of amplitudes and phases of the effect. Another observation worth noting is that the corrected AMs of the independent frequencies follow less closely to that of $f_\mathrm{orb}$ than the raw AMs (as revealed by a careful comparison of the two panels of Figure\,\ref{sdb2stable}).

Figure\,\ref{campm} compares AMs of $f_\mathrm{orb}$ and its first harmonic with the {other} three frequencies, {$f_{02}$, $f_{12}$ and $f_{15}$}, with available measurements in KIC\,2438324. We find that the modulation patterns are {significantly} different in these cases, in particular for $f_{12}$ whose AM exhibits a large $\sim50\,\%$ peak-to-peak variation. This frequency seems {to} experience three local maxima and minima (with two missing measurements due to the amplitude becoming too low). Moreover, we note that the two other frequencies, $f_{02}$ and $f_{15}$, have AMs that are not following the general trend of $f_\mathrm{orb}$ in general, although there is some overlap between the modulations.

Figure\,\ref{sdb1com} shows the comparison of AMs between the orbital frequency, $f_\mathrm{orb}$, and 11 independent frequencies of KIC\,11179657. We only provide the raw AM patterns for these frequencies considering that this star is only slightly contaminated (also see Figure\,\ref{triplet2} and \ref{doublet2}). In contrast with what we observe for KIC\,2438324, we find that $f_{03}$ presents a very similar modulation pattern compared to $f_\mathrm{orb}$ over the entire observation run. However, if we compare the modulation patterns in each observational segment, then almost all frequencies show similar AM patterns in at least one segment, such as $f_{02}$ (in the first part) and $f_{05}$ (in the middle and last parts), considering the uncertainties of the measurements. In addition, we find that the two independent frequencies may have similar AMs {at some observational segment. For instance}, AM patterns of $f_{09}$  and $f_{12}$ are found to be somewhat similar to the AM of $f_{02}$ in the middle observation segment. Whether {this is a coincidence or due to calibration issues} needs further investigation.

\section{Discussion}
The results presented in this paper are the continuation of our project to analyze AMs and FMs in compact pulsators by exploiting the full data sets available from {\sl Kepler}. A natural explanation for such phenomena is the weak nonlinear interaction between resonant modes, which can produce diverse modulation patterns \citep{buchler1995,buchler1997}. However, our detailed studies of KIC\,2438324 and KIC\,11179657 lead to the discovery of systematic AMs for several of the independent pulsation modes identified in these two stars. This  behavior is different from previous modulations characterized in several compact pulsating stars \citep[][Paper\,\RNum{1}]{zong2016a,zong2016b}. It is plausible that this kind of AMs may be induced by instrumental effects onboard or by the data reduction pipeline. This finding adds further complexity to the goal of precisely measuring the intrinsic amplitudes of oscillations or other signals (e.g., transits of planets).

\subsection{Comparison of frequency content with previous results}
KIC\,2438324 and KIC\,11179657 show a much lower number of frequencies than that detected in KIC\,3527751 (Paper\,\RNum{1}), i.e., a few tens compared to more than two hundred. In both stars, the frequency contents are similar with modes detected in the [100, 600]\,$\mu$Hz range. In KIC\,2438324, most of the frequencies already detected in \citet{2011ApJ...740L..47P} are recovered, except one low-amplitude (suspected) frequency at 405.45\,$\mu$Hz although we have analyzed a light curve more than six times longer than theirs. From our analysis, we only have four additional low-amplitude frequencies not seen in \citet{2011ApJ...740L..47P}, which complete the triplet near 319\,$\mu$Hz and the independent frequency around 290\,$\mu$Hz that is now a doublet. Compared to \citet{2012MNRAS.422.1343P}, we have recovered all their frequencies from our 5.6$\sigma$ level, except one at 262.8\,$\mu$Hz which is nonetheless obtained if we allow $S/N\sim 5.1$ to be a trustworthy detection. As the noise level decreases, we find eight additional weak frequencies, which complete the former doublet at 195\,$\mu$Hz to make it a triplet, and two former independent frequencies near 206 and 369\,$\mu$Hz which are now doublets. 

Both stars show a rich set of multiplets interpreted as rotational splittings and relatively few independent frequencies. These may prove to be good candicates for further seismic modelling \citep[see, e.g.,][]{2005A&A...437..575C}. At present, there is no detailed seismic result obtained for sdB pulsators with the full {\sl Kepler} photometry yet. Considering their similar frequency contents and orbital periods, it will be interesting to compare their internal structural and dynamical properties through the technique of asteroseismology \citep[see, e.g.,][]{2008A&A...489..377C}.

\subsection{Modulation patterns}
\begin{figure*}
\centering
\includegraphics[width=8.2cm]{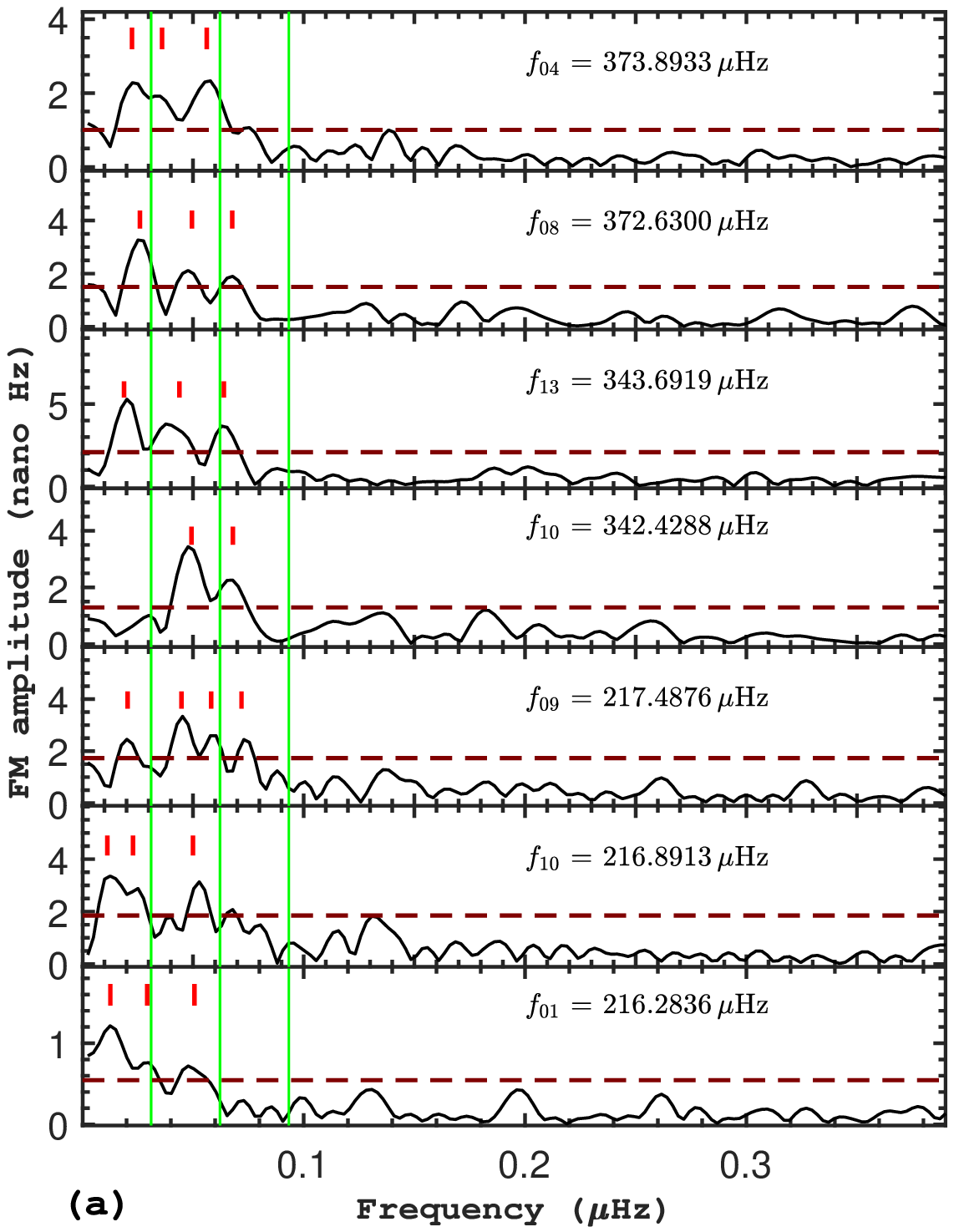}
\hspace{0.28cm}
\includegraphics[width=8.2cm]{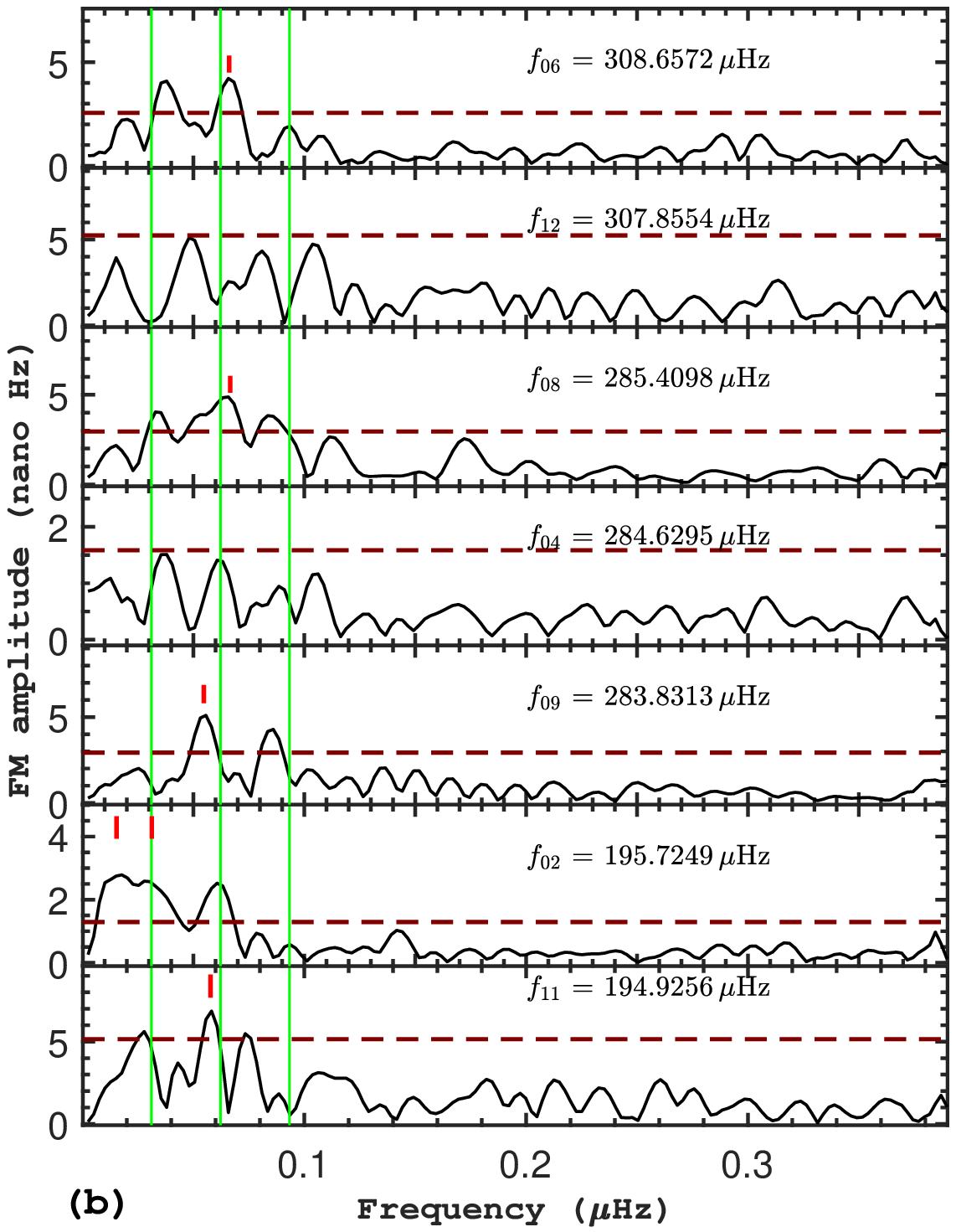}
\caption{The LSPs of frequency modulations concentrating on the representative frequencies in KIC\,2438324 (a) and KIC\,11179657 (b). The Vertical 
line indicates the {\sl Kepler} orbital frequency, $f_{K} = 0.0311\,\mu$Hz, and its harmonics. The dashed horizontal line represents the typical 
4\,$\sigma$ level of local noise. The detected frequencies above that level are found with the values where the vertical segments stand. Note that 
LSPs in (b) are resolved with sidelobes due to observation gaps.
\label{fmam}}
\end{figure*}
%==========================================================================

All representative pulsation modes illustrated in this paper are discovered with variations in amplitude and frequency. Considering that our evaluations of the uncertainties are proven to be robust (Section\,\ref{s:r}), we believe that the values determined for the amplitudes and frequencies are highly confident. Besides, AMs and FMs from the two different flux calibrations (raw and corrected) can serve as a way to double check the determined values. Our results suggest that the frequency modulations show consistent patterns, when measured alternatively from the raw and corrected fluxes, for both stars (see Figure\,\ref{triplet}-\ref{doublet2}). However, the correction of the flux introduces a significant difference in AMs patterns when the fraction of contaminating light coming from nearby background or foreground objects is large. For instance, in KIC\,2438324 we not only observe that the values of peak-to-peak variation are different, but also that the phase of the AMs shifts between raw and corrected data for the same frequency. The corrected AMs {also} follow that of $f_\mathrm{orb}$ less closely than the raw AMs. All these observations suggest that the correction of raw flux brings an extra signal to AMs, as the contamination factor increases. FMs remain unaffected by this effect because we extract the highest peaks even if the AMs induce symmetric (weak) side-bands around those peaks in LSP.

We recall that nonlinear resonant interactions between three modes forming a triplet can produce various AMs and FMs \citep{buchler1995,buchler1997,goupil98} featuring stable, periodic, or irregular modulation pattern. All the representative frequencies discussed previously satisfy this triplet resonance condition (doublets are incomplete triplets with an undetected component). Considering the uncertainties, the detected AMs and FMs in the representative frequencies might be associated with the intermediate regimes of the triplet resonance \citep[see such examples in][]{zong2016a,zong2016b}. 
We clearly see quasi-periodic FMs occurring on timescales of several months to {well over a year}, as shown in Figure\,\ref{fmam}, which is comparable to the timescale that corresponds to the frequency mismatch of $\sim0.01\,\mu$Hz and $\sim0.02\,\mu$Hz in the triplets 217\,$\mu$Hz (KIC\,2438324) and 284\,$\mu$Hz (KIC\,11179657), respectively. From this figure, we clearly see various configurations, which indicates that their modulation patterns are indeed different (as roughly displayed in Figure\,\ref{triplet}-\ref{doublet2}). Nevertheless, as nonlinear resonant coupling theory predicts, similar timescales of FMs within the same multiplet are observed. 
Several resonant components, whose frequencies evolve in phase or anti-phase during the observations, show similar modulation patterns to those found between the components of the quintuplet $Q_1$ in KIC\,3527751 (Paper\,\RNum{1}). We note that most of the FM timescales are not close to that of {\sl Kepler}'s orbit and its resonances which was recently reported on phase modulations in non-Blazhko RR Lyrae stars \citep{2019MNRAS.485.5897B}.
However, the measured AMs, particularly in KIC\,2438324, are contaminated by a systematic modulation pattern (discussed later) and by the correction of the light fraction associated to the target. This pollution can seriously impair the recovery of intrinsic amplitude modulation patterns of oscillation modes, which is the most straightforward way to characterize nonlinear effects on stellar pulsations. These systematic patterns will need to be removed prior to any detailed quantitative comparison with theoretical predictions. Our suggestion, in the meantime, is that stars without contaminating light should be given a higher priority for such studies. Finally, we point out that the modulations may become more complicated as the number of the detected frequency increases, comparing AMs and FMs detected in KIC\,2438324, KIC\,11179657, KIC\,10139564 and KIC\,3527751.

\subsection{Instrumental effects?}
%=++++++++++++++++++++++++++++++++++++++++++++++++++++++++++++++++++++++++
\begin{figure}
\centering
\includegraphics[width=8cm]{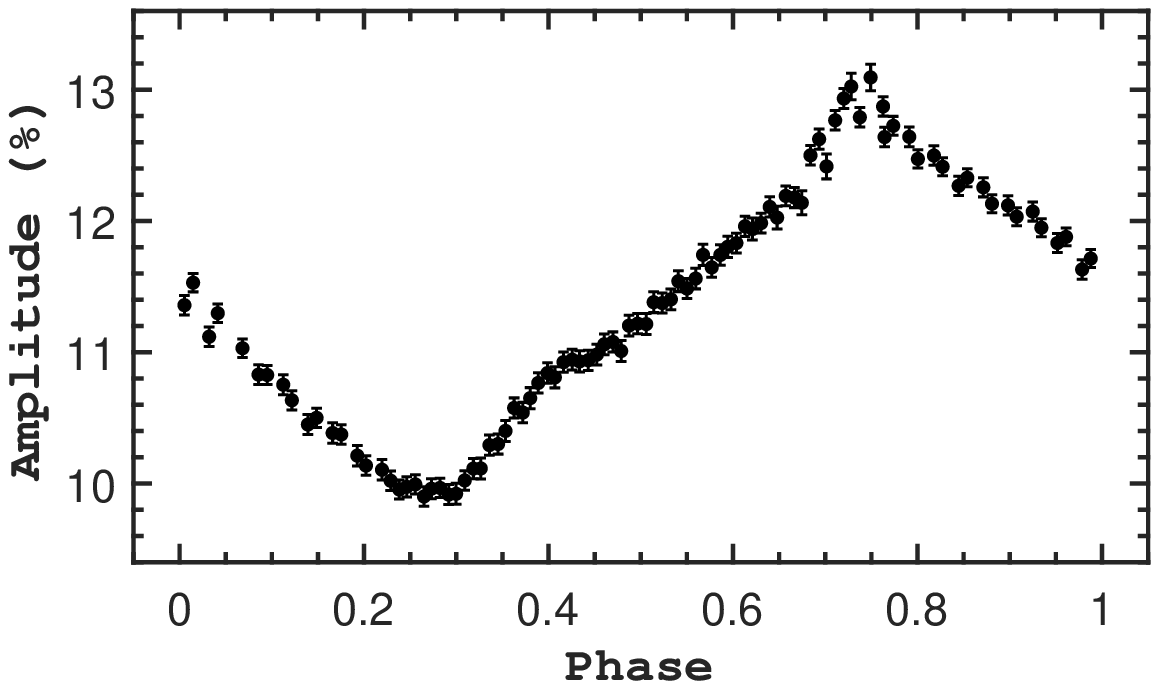}\\
\vspace{0.2cm}
\includegraphics[width=8cm]{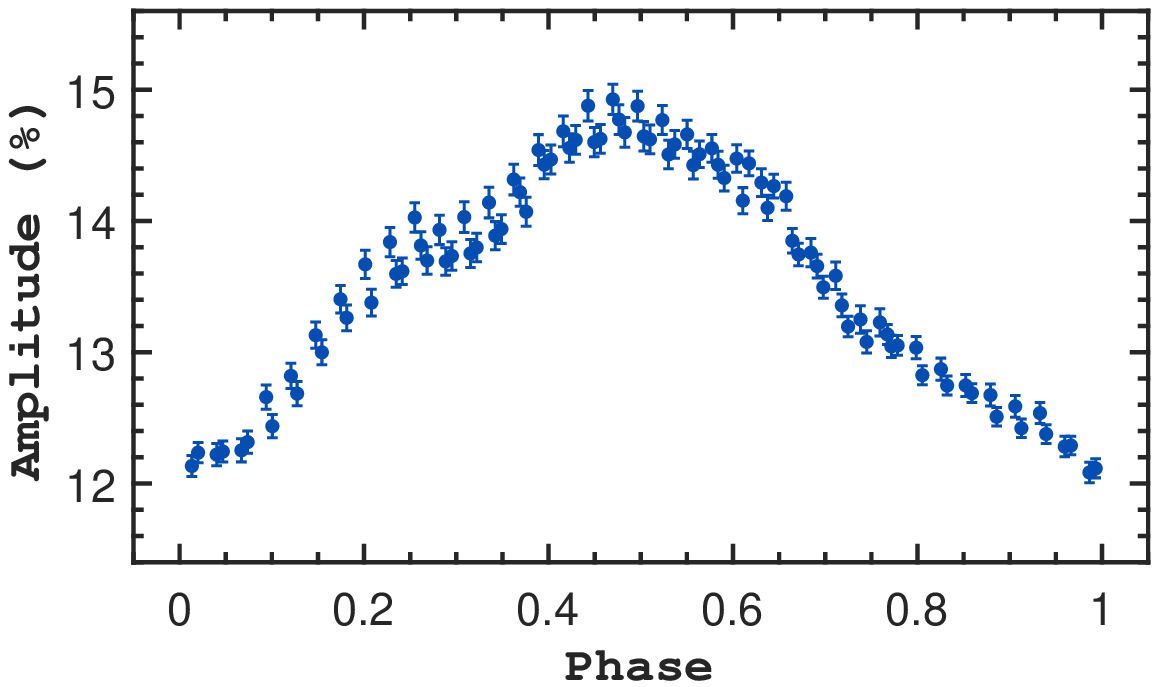}
\caption{Phase diagram of the amplitude variation of the orbital frequency in KIC\,2438324: ({\sl Top} panel) raw flux and 
({\sl bottom} panel) corrected flux.
\label{phase}}
\end{figure}
%==========================================================================

At odds with what was found for the star KIC\,3527751, we discovered systematic AM patterns in several independent modes and in the orbital signals of the two stars considered in this paper. The orbital signal, expected to be constant in amplitude, was used as a reference to calibrate intrinsic modulation patterns. This is the first time that this systematic is characterized to better measure AMs and FMs of oscillation modes in pulsating compact stars. In previous studies, we had not encountered this kind of modulation pattern although we had processed more than 300 frequencies \citep[see][and Paper\,\RNum{1}]{zong2016a,zong2016b}. Systematic AMs are likely different from star to star, but they can be distinguished from the various AMs induced by nonlinear mode interactions. A key feature on these systematic modulations is that they have a timescale identical to {\sl Kepler}'s orbital period, $P\sim372.5$\,days. Figure\,\ref{phase} shows a phase diagram constructed on the 372.5\,d period for the AM of $f_\mathrm{orb}$ in KIC\,2438324. The folded curves clearly show that the right period was used for the raw and corrected AMs (with different zero-points in phase). To confirm the period, we used the PDM (phase dispersion minimization) to evaluate the period, which gives a value of 372.7\,d {and 371.5\,d} for the raw {and the corrected} AM, {respectively}. Similarly, a value of 373.1\,d was obtained from a Lomb-Scargle periodogram of the raw AM curve. This modulation pattern of timescale identical to the orbital period of {\sl Kepler} cannot be coincidental and is most likely induced by instrumental effects onboard. 

This finding shows that one should be cautious when exploiting {\sl Kepler}'s light curves to measure long-term amplitude modulations, particularly with periods near to the spacecraft orbit. For instance, a similar small-amplitude fluctuation was uncovered in M-giant stars based on {\sl Kepler}'s photometry \citep{2013MNRAS.436.1576B}, with several stars showing a systematic brightness variation on a {\sl Kepler}-year timescale of unclear origin. Such variations can be harmful for many science objectives, such as determining the transit depth for wide-orbit exoplanets, which can be used to measure oblateness of transiting planets \citep[see, e.g.,][]{2017AJ....154..164B}. As the spacecraft rolls by 90 degrees every season, the fraction of contaminating light changes every quarter, which could possibly be related to the {\sl Kepler}-year periodic AMs seen in KIC\,2438324. However, the AMs are still showing a {\sl Kepler}-year period (Figure\,\ref{phase}) in the corrected fluxes, besides the fact that these have been rectified by the contamination factor. This indicates that the contamination factor may not be estimated precisely enough within and across quarters. In addition, quantum efficiency could also differ when stellar light is acquired from different pixels. Moreover, the point spread function of the stars will also change as the telescope roll angle changes every quarter. 
Further investigation will need to be carried out with target pixel photometry, which may help us to remove this systematic amplitude variations. A plausible reason might be that the neighbour pixels around that target still contain a very small fraction of the flux \citep[see, e.g.,][]{2017A&A...598A..74P}, meaning 
that we do not count the right amount of flux for the target when using the standard {\sl Kepler}'s reduction pipeline. We finally recall that this kind of systematics in AMs needs to be given careful inspection also for the ongoing mission {\sl TESS} \citep{2014SPIE.9143E..20R}, especially for targets observed over several sectors.

\subsection{General prospectives}
Space missions, such as {\sl Kepler} \citep{2010Sci...327..977B} and {\sl TESS} \citep{2014SPIE.9143E..20R}, bring opportunities to measure frequencies in pulsating stars with accuracy of tens of micro Hz down to a few nano Hz. We are presently capable of measuring frequency variations on very tiny scales, allowing us to study new physics, such as how to precisely calculate FM patterns induced by nonlinear effects on pulsation modes. {In particular,} progress in this direction would allow to securely determine secular rates of frequency variations due to stellar evolution, or to detect with a higher confidence the presence of small orbiting objects, with residual modulations after {the} nonlinear FM correctly removed. This has received particular interest in compact pulsators these past decades, with the possibility to constrain the rate of pulsation period change \citep{kepler2005,vauclair11,hermes2013} and uncover periodic phase variations caused by the presence of planetary companions \citep{silvotti2007,2018A&A...611A..85S}. In the context of the two close binary systems KIC\,2438324 and KIC\,11179657 discussed in this paper, after correcting for the complex FMs due to nonlinear interactions, the residual modulation could also be used to measure the rate of spin-up induced by tides from the companion, if this acceleration is rapid enough, or set upper limits otherwise.

Even though flux can be measured with high precision from space instruments, consecutive long-term observations are not free of systematic uncertainties that can affect such measurements. We have shown that thorough inspections of the obtained amplitudes must be carried out if the goal is to characterize accurately amplitude variations over long timescales \citep[to study, e.g., physics of nonlinear mode interactions inside stars in][]{goupil1994,buchler1995}. In the nonlinear physics context, any instrumental bias to the intrinsic values of AMs would lead to results far away from the real modulation patterns, which will likely affect the critical physical quantities from the calculations of amplitude equations.  

\section{Conclusion}
In this paper, the second of a series devoted to a systematic survey of pulsations in evolved compact stars (hot B subdwarf and white dwarf stars) as observed from the {\sl Kepler} spacecraft, we focus on the comparison of AMs and FMs of oscillation modes as measured from the two types of flux calibrations, raw and corrected, delivered by the standard data reduction pipeline. This is done for two pulsating sdB stars, KIC\,2438324 and KIC\,11179657, which are both primary components of a binary system. The first goal was to precisely measure the intrinsic AMs and FMs of oscillations in those compact stars, further paving the way to theoretical and quantitative calculations within the framework of nonlinear stellar oscillation theory that could rely on such observations.

We first extended our estimation of uncertainties in measuring amplitudes and frequencies by testing 50\,000 artificial signals, a large improvement over the more limited set of 1000 signals originally used by \citet{zong2016b}. These new simulations agree well with the previous results of \citet{zong2016b}, and demonstrates amplitude and frequency errors all are measured accurately, independent of the $S/N$ ratio of the signal. However, uncertainties in the phase determination may be overestimated depending on $S/N$ ratio (Figure\,\ref{simulation}). The latter, however, are not used for our objectives, yet. With these quantitative tests, we could precisely and confidently extract frequencies from the light curves of the two sdB stars using the two different flux calibrations, raw (SAP) and corrected (PDC-SAP). Since the photometry is longer than that in previous published literature about these objects, we were able to resolve 22 and 34 frequencies in KIC\,2438324 (Table\,\ref{t1}) and KIC\,11179657 (Table\,\ref{t2}), respectively. These bring {an} additional four and eight low amplitude modes to the formerly available lists of detected pulsations in these stars. We mention that both stars show very similar and relative simple frequency contents, and could be of high interest for future comparative seismic analyses. 

We then precisely measured AMs/FMs for frequencies with amplitude down to $\sim0.7$\,ppt and $\sim0.3$\,ppt in KIC\,2438324 and KIC\,11179657, respectively. By comparing modulation patterns measured from two different kinds of flux, we find that AM patterns change significantly for the modes observed in KIC\,2438324, which suffer from significant light contamination by nearby stars (Figure\,\ref{sdb2stable}), but not for the modes in KIC\,11179657 whose contamination factor is almost zero (Figure\,\ref{sdb1com}). Differing from our previous studies, we identify clear systematic AMs of frequencies in KIC\,2438324 through inspection on many independent modes and binary signals. Several methods have been used to determine the timescale of this regular AM, leading to a periodicity of about 372.5\,days which is identical to {\sl Kepler}'s orbital period around the Sun (Figure\,\ref{phase}). We argue that this systematic AM is an additional difficulty, which to some extent can impair an accurate determination of the intrinsic AMs needed to constrain theoretical calculations of nonlinear couplings of pulsation modes, but also, more generally, for the study of long-period variables or transiting exoplanets. However, we stress that this effect can be corrected provided that some reference of normally constant amplitude exist, which is the case for pulsating stars in binary systems. In {summary}, these findings suggest that stars without contamination could be much easier to rectify the systematic AMs because the correction of {light} pollution could destroy the real AM patterns.

Although the AMs evolve following different patterns in KIC\,2438324, the FM patterns are found to be unaffected {from these two independent measurements with different} {types} of flux. All the representative frequencies exhibit frequency variations in a relative narrow scale (Figure\,\ref{triplet}-\ref{doublet2}), typically of 10-20 nano~Hz, which is of the same order as the frequency mismatch found in the triplets at 217\,$\mu$Hz in KIC\,2438324 (Figure\,\ref{triplet}) and 284\,$\mu$Hz in KIC\,11179657 (Figure\,\ref{triplet2}). These FMs show somewhat (anti-)correlations between different components within the same multiplet, with peak-to-peak periodicities ranging from months to years. This suggests that nonlinear weak mode interaction happens between the involved components, as expected from the nonlinear mode coupling theory \citep[e.g.,][]{buchler1995}.
We recall that similar results have been obtained in our previous studies, such as in KIC\,3527751 (Paper\,\RNum{1}). Observation of these FMs are of particular interest for future comparisons with nonlinear calculations using, e.g., the amplitude equation formalism \citep{goupil1994}.

In forthcoming steps, we will concentrate on the other pulsating sdB stars with appropriate data available for such kind of study in order to provide a more general statistical view on these modulations. We stress that the comparison described in this paper also applies to the photometry gathered from the {\sl TESS} mission, since these data are processed through a very similar pipeline to the one used for {\sl Kepler} \citep{2016SPIE.9913E..3EJ}.
We expect that AMs/FMs of oscillation modes in compact pulsators from {\sl Kepler} observations, as well as from the ongoing {\sl TESS} monitoring \citep{charpinet2019}, will ultimately provide a solid base for future theoretical investigations of nonlinear effects in stellar pulsations which may lead to new insight in the physics of stars and their oscillations, through, e.g., the determination of linear growth rates of modes.  

\facility{\sl Kepler}
\software{\felix{} \citep{charpinet2010}}

\acknowledgments
W.Z. acknowledges support from the National Natural Science Foundation of China (NSFC) through the grant 11903005 and 11833002, and the support from the Fundamental Research Funds for the Central Universities. S.C. and G.V. acknowledge support from the Agence Nationale de la Recherche (ANR, France) under grant ANR-17-CE31-0018, funding the INSIDE project, and financial support from the Centre National d'Études Spatiales (CNES, France). The authors gratefully acknowledge the {\sl Kepler} team and all who have contributed to making this mission possible. Funding for the {\sl Kepler} mission is provided by NASA's Science Mission Directorate.

%\clearpage
\bibliographystyle{aasjournal}



\end{document}